\documentclass[twocolumn]{aastex702}
\usepackage{amsmath,amssymb,mathrsfs}

\newcommand{\nthp}{N$_2$H$^+$}
\newcommand{\hcop}{HCO$^+$}
\newcommand{\kms}{km\,s$^{-1}$}

\newcommand{\twco}{$^{12}$CO}
\newcommand{\cc}{cm$^{-3}$}
\newcommand{\sqc}{cm$^{-2}$}
\newcommand{\Jypbm}{Jy\,beam$^{-1}$}
\newcommand{\Itr}{$I$(3\,mm)}
\newcommand{\Ione}{$I$(1\,mm)}

\newcommand{\naoj}{National Astronomical Observatory of Japan, 2-21-1 Osawa, Mitaka, Tokyo 181-8588, Japan}
\newcommand{\shiz}{Faculty of Global Interdisciplinary Science and Innovation, Shizuoka University, 836 Ohya, Suruga-ku, Shizuoka 422-8529, Japan}
\newcommand{\sokendai}{Astronomical Science Program, Graduate Institute for Advanced Studies, SOKENDAI, 2-21-1 Osawa, Mitaka, Tokyo 181-1855, Japan}

\newcommand{\shibaura}{Materials Science and Engineering, College of Engineering, Shibaura Institute of Technology, 3-7-5 Toyosu, Koto-ku, Tokyo 135-8548, Japan}
\newcommand{\stony}{Stony Brook University, Stony Brook, NY, 11794-3800, USA}

\newcommand{\oan}{Observatorio Astron\'omico Nacional (IGN), Alfonso XII 3, 28014, Madrid, Spain}

\newcommand{\mpia}{Max-Planck-Institut f\"ur Astronomie, K\"onigstuhl 17, 69117, Heidelberg, Germany}
\newcommand{\mpifr}{Max Planck Institute for Radio Astronomy, Auf dem H\"ugel 69, 53121 Bonn, Germany}

\newcommand{\jao}{Joint ALMA Observatory, Alonso de C\'ordova 3107, Vitacura, Santiago 763-0355, Chile}

\newcommand{\ipas}{Faculty of Pure and Applied Sciences, University of Tsukuba, 1-1-1 Tennodai, Tsukuba, Ibaraki, 305-8577, Japan}

\newcommand{\tomonaga}{Tomonaga Center for the History of the Universe, University of Tsukuba, 1-1-1 Tennodai, Tsukuba, Ibaraki, 305-8577, Japan}

\newcommand{\tiar}{Tsukuba Institute for Advanced Research (TIAR), University of Tsukuba, 1-1-1 Tennodai, Tsukuba, Ibaraki, 305-8577, Japan}

\newcommand{\ioa}{Institute of Astronomy, The University of Tokyo, 2-21-1, Osawa, Mitaka, Tokyo 181-0015, Japan}

\newcommand{\iasku}{The Institute for Advanced Study, Kyushu University, Nishi-ku, Fukuoka 819-0395, Japan}

\newcommand{\epsku}{Department of Earth and Planetary Sciences, Faculty of Science, Kyushu University, Nishi-ku, Fukuoka 819-0395, Japan}
\graphicspath{{./}{figures/}}

\begin{document}

\title{ALMA observations towards M83 reveal \nthp\ tracing dense gas}

\author[0000-0002-6824-6627]{Nanase Harada}
\email{nanaseharada@gmail.com}  
\affiliation{\naoj}
\affiliation{\sokendai}

\author[0000-0003-0563-067X]{Yuri Nishimura}
\email{nishimura.yuri.ke@u.tsukuba.ac.jp}  
\affiliation{\ipas}
\affiliation{\tomonaga}
\affiliation{\tiar}
\affiliation{\ioa}

\author[0000-0002-2501-9328]{Toshiki Saito}
\email{saito.toshiki@shizuoka.ac.jp}  
\affiliation{\shiz}

\author[0000-0002-2569-1253]{Mario Tafalla}
\email{m.tafalla@oan.es}  
\affiliation{\oan}

\author[0000-0002-8762-7863]{Jin Koda}
\email{jin.koda@stonybrook.edu}  
\affiliation{\stony}

\author[0000-0002-1959-7201]{Doris Arzoumanian}
\email{arzoumanian.doris.958@m.kyushu-u.ac.jp}  
\affiliation{\iasku}
\affiliation{\epsku}

\author[0000-0002-9333-387X]{Sophia K. Stuber}
\email{astro@sophiastuber.de}  
\affiliation{\naoj}
\affiliation{\mpia}
\affiliation{\mpifr}

\author[0000-0002-0465-5421]{Akihiko Hirota}
\email{akihiko.hirota@alma.cl}  
\affiliation{\jao}
\affiliation{\naoj}

\author[0000-0002-9668-3592]{Yoshimasa Watanabe}
\email{nabe@shibaura-it.ac.jp}  
\affiliation{\shibaura}



\begin{abstract}
Galactic observations have shown that \nthp\ intensities preferentially trace star-forming dense molecular gas compared with HCN or \hcop, which are conventionally used as dense-gas tracers in extragalactic studies. In this study, we aim to verify this trend in the nearby star-forming galaxy M83. Using ALMA, we observed the $J=1$--$0$ transitions of \nthp\ and HCN, along with the 1-mm and 3-mm continuum, and analyzed the archival CO(1--0) data. The data achieved a spatial resolution of 48 pc in the southwestern bar-end region of M83, covering an area of $2.1$\,kpc$\times1.8\,$kpc. \nthp\ and HCN intensities are approximately linearly correlated, indicating that HCN luminosity can still reasonably predict the amount of dense gas as well as \nthp\ luminosity does. However, in regions with high column densities, \nthp\ becomes a super-linear function of HCN, suggesting that the \nthp\ emission is more sensitive to dense gas than HCN. We employed a radiative transfer model that predicts line intensities as a function of column density, adopting parametric column density probability distribution functions (N-PDFs) to describe the underlying column density distribution. The model demonstrates that the observed linear or superlinear trend can be explained by a log-normal distribution only or a log-normal and a power-law tail distribution in the N-PDF. Furthermore, the \nthp\ intensity is significantly more sensitive to the power-law tail than the HCN intensity. In addition, our analysis incorporating archival data shows that \nthp\ is a less reliable tracer of dense gas in galactic center regions, where warmer temperatures may promote \nthp\ destruction.
\end{abstract}

\keywords{}


\section{Introduction} \label{sec:intro}
Within molecular clouds, there is a hierarchy of density structures. The densest parts are directly associated with star formation, as they are most favorable for gravitational collapse. It has been suggested that there is a density threshold for star formation. Some observational studies proposed a specific volume density ($n\sim10^4$\,\cc) or column density \citep[$N\sim10^{22}$\,\sqc;][]{2010ApJ...724..687L}, whereas there are theories that consider the threshold as a function of the mean density, Mach number, and the virial parameter \citep{2005ApJ...630..250K}. More recently, the star formation threshold has been shown to be linked to the minimum line mass of molecular filaments, which fragment to dense star-forming cores \citep{2014prpl.conf...27A,2023ASPC..534..233P}. While it is possible to resolve these dense cores, filaments, and clumps in Galactic star-forming regions, it is unfeasible in extragalactic observations except in the most nearby galaxies using currently available observing capabilities whose beam sizes are typically on the order of 10\,pc or more. Instead of directly resolving these dense cores, certain molecular transitions with high critical densities can be useful in measuring the amount of dense gas in extragalactic sources. 

For example, the HCN($J=$1--0, hereafter 1--0) transition has a critical density of $n_\mathrm{crit}\sim 10^6$\,cm$^{-3}$ at 10\,K in the optically thin case with the two-level approximation. \citet{2004ApJS..152...63G} found that it correlates more directly with the star formation rate (with the power law of $\sim1$) than CO(1--0) does (with the power law of 1.25). Using HCN as a dense gas tracer, many studies observed various locations within galaxies to study the link between the dense gas content and star formation \citep[e.g., ][]{2016ApJ...822L..26B,2018ApJ...858...90G,2019ApJ...880..127J,2023MNRAS.521.3348N}. 

A theoretical investigation by \citet{2007ApJ...669..289K} considered the density distribution within molecular clouds and performed radiative transfer modeling of CO, HCN, and HCO$^+$ (1--0), which reproduced the power law of $\sim1.5$ in SFR as a function of CO luminosity while the power becomes close to unity as a function of HCN. This study successfully explained the super-linear relationship between the star-formation rate and the CO luminosity, as well as the linear relationship between the SFR and the HCN luminosity.

However, Galactic observations on the molecular cloud scale started to challenge this picture. Studies in the parsec- to 10-pc scale observations consistently revealed that the HCN(1--0) and \hcop(1--0) transitions are relatively extended, and come from not only dense ($n\gtrsim 10^4$ cm$^{-3}$), but also from translucent ($n\sim 10^3$ cm$^{-3}$) gas \citep{2017A&A...605L...5K,2017ApJ...848...17N,2017A&A...599A..98P,2017ApJ...845..116W,2020MNRAS.497.1972B,2021A&A...646A..97T,2023A&A...679A.112T}. 

These studies have shown that \nthp\ is a more selective tracer of dense gas than HCN in the $J=$1--0 transition. Although its critical density is lower ($n_\mathrm{crit}=10^5$\,cm$^{-3}$) than that of HCN ($n_\mathrm{crit}=10^6$\,cm$^{-3}$)\footnote{These values are calculated for an optically thin case at $T=10\,$K. We also note that HCN is often optically thick, and the critical density is likely lower \citep[e.g., ][]{2015PASP..127..299S}.}, its enhanced abundance in cold and dense gas makes it emit mostly in dense regions while HCN can also emit in translucent regions. Galactic observations have already used \nthp\ to study high-density filaments \citep[e.g., ][]{2014ApJ...797...76L}. Theoretical studies have also confirmed that this transition predominantly emits from higher-density regions \citep{2019ApJ...871..238H,2023MNRAS.524.5971P,2023MNRAS.526.4952P,2024MNRAS.531.4408P}.

In response to these Galactic findings, extragalactic observers have begun to explore the behavior of the \nthp\ intensities in various environments. Although this transition is weaker than that of HCN, it is possible to detect \nthp(1--0) in the central regions of bright galaxies with ALMA or NOEMA with a reasonable observing time \citep{2019ApJ...884..100H,2022A&A...659A.173E}. Some studies have mapped the \nthp(1--0) transition as a part of line surveys \citep{2021A&A...656A..46M,2023ApJ...955...27N,2024ApJS..271...38H}.

On the other hand, detection in disk regions requires more observing time. Yet, it is important to investigate the emission in these regions since they host less extreme physical conditions than galactic centers, which are characterized by high cosmic-ray ionization rates, high pressure, and high temperatures. \citet{2014ApJ...788....4W} conducted a line survey towards two spiral arm regions with different star formation rates in M51, and detected a number of species including \nthp. \citet{2023A&A...676L..11J} observed \nthp\ and HCN $J$=1--0 transitions towards five selected positions in a double-barred spiral galaxy NGC 6946 using the IRAM 30-meter telescope. This study found that the \nthp/HCN intensity ratio is constant among these positions on the kpc scale. If the \nthp/HCN ratio is constant in different physical conditions and across spatial scales, then it would be reasonable to use HCN emission as a tracer of the dense gas instead of the fainter \nthp\ line. However, this ratio starts to deviate from a constant at a higher spatial resolution. The SWAN survey (Surveying the Whirlpool at Arcseconds with NOEMA) observed towards the central 5 kpc $\times$ 7 kpc region of M51 with a 125-pc resolution using NOEMA \citep{2023A&A...680L..20S,2025A&A...696A.182S,2025arXiv250719439S}, and found that the \nthp\ and HCN intensities are superlinearly related in the disk region. Because it is in general true that the higher column density regions have higher volume densities, this result suggests that \nthp\ emits more efficiently at higher volume density regions.


While some studies have shown that \nthp\ may be a better tracer of dense gas than the more commonly used HCN, the reliability of \nthp\ to trace dense gas is still not fully explored in extragalactic sources. In this paper, we present \nthp(1--0) observations with a $48\,$pc resolution toward the bar-end region of the barred spiral galaxy M83 using the Atacama Large Millimeter/submillimeter Array (ALMA). We chose the observation area in the disk region because the galactic center is harder to interpret due to its complexity mentioned earlier. We also compare these observations with HCN(1--0) and the 1-mm and 3-mm continuum observations. The continuum at the 1-mm wavelength mostly traces dust, while it originates from star formation (free-free emission) at 3\,mm. This study achieves higher angular resolution than the study toward M51 by \citet{2023A&A...680L..20S}, although the field of view is smaller. This paper is structured as follows. In Section \ref{sec:obs}, we describe our observations. We present our observational results in Section \ref{sec:results}, and radiative transfer modeling to interpret the observational results in Section \ref{sec:rad}, while the implications of our results are discussed in Section \ref{sec:disc}. Finally, our main conclusions are summarized in Section \ref{sec:summary}.

\section{Observations and data reduction} \label{sec:obs}
We target the nearby barred-spiral galaxy M83 \citep[$d=4.5$\,Mpc;][]{2003ApJ...590..256T}. Because of its face-on configuration \citep[$i=26^\circ$;][]{2023ApJ...949..108K} and proximity, it is an ideal target for studying star formation in galaxies. There have been studies targeting M83 covering multiple wavelengths from cm-wave radio continuum \citep{2020MNRAS.495..479R}, HI \citep{2023A&A...675A..37E}, mid- and far-IR \citep{2012MNRAS.421.2917F}, near-IR \citep{2025arXiv250508874K}, optical \citep{2022A&A...666A..29D}, UV \citep{2005ApJ...619L..83B}, and X-ray \citep{2016ApJ...824..107Y}. There have also been extensive CO observations using ALMA \citep{2023ApJ...949..108K,2025arXiv250508876K}.

In this paper, we present $J$=1--0 transitions of HCN, \nthp\, and 1-mm ($\sim$349\,GHz) and 3-mm ($\sim$96\,GHz) continuum data from three frequency settings taken by ALMA (project code 2021.1.01195.S; PI: Harada). Two of them used the Band 3 receiver at the 3-mm wavelength to observe $J=$1--0 transitions of the species pairs HCN-\hcop\ and HNC-\nthp. A frequency setting using Band 7 at the 1-mm wavelength is designed to target the dust continuum while it contains $J=$4--3 transitions of HCN and \hcop\ in the band. Neither of these $J=$4--3 transitions is detected. \citet{2024ApJ...969...82H} already presented the HCN, \hcop, and HNC (1--0) transitions. We also use the CO(1--0) data presented in \citet{2023ApJ...949..108K} for the analysis in this paper. Because all the transitions discussed in this paper are in the $J=$1--0, we may not specifically write the quantum numbers of the transitions.

The observation consisted of a mosaic that covered a rectangular region of $94\arcsec \times 80 \arcsec$ ($2.1$\,kpc$ \times 1.8\,$kpc) with a position angle of 78 degrees (see Figure \ref{fig:mom0}f). This region was observed using the 12-meter, 7-meter, and total power (TP) arrays.
The interferometer data (12-meter and 7-meter antennae) were imaged with the PHANGS pipeline \citep[ver. 2;][]{2021ApJS..255...19L} using CASA \citep{2007ASPC..376..127M, 2022PASP..134k4501C} version 5.6.3.
For further analysis, we used CASA version 6.6.0. For the total power data, we used the products obtained from the ALMA data delivery (QA2), and the CASA task \texttt{feather} to combine all the arrays.
All products were convolved into a common angular resolution of $2\farcs2 \times 2\farcs2$ (=48\,pc $\times$ 48\,pc) to match the largest beam, which is the CO resolution in this case. The products have single-channel sensitivities of 30 mK (HCN) and 11 mK (\nthp) over 1\,MHz ($\sim$3\,\kms). We applied primary beam correction to all the cubes. The 3-mm and 1-mm continua have rms of $10\,\mu$\Jypbm\ and $250\,\mu$\Jypbm, respectively. We note that ALMA TP observations cannot provide a continuum map, so these data are subject to missing flux. Because HCN emission with and without TP shows a difference less than 10\% \citep{2024ApJ...969...82H}, we consider that the missing flux does not change our results too much for the 3-mm continuum. As for the 1-mm continuum, the maximum recoverable scale of the 7-m array observation is about $19''$, which should cover most of the structure. However, we cannot deny the possibility that it may miss some of the diffuse gas traced by CO. 

In addition to these observations, we also used the archival data of \nthp(1--0) and HCN(1--0) in the central region of M83 presented in \citet{2019ApJ...884..100H} and \citet{2021MNRAS.505.4310C}, and reduced according to the procedure presented by  \citet{2024ApJ...969...82H}.

\section{Observational Results}\label{sec:results}
\subsection{Velocity-integrated and continuum images}\label{sec:images}

\begin{figure*}
\includegraphics[width=0.99\textwidth]{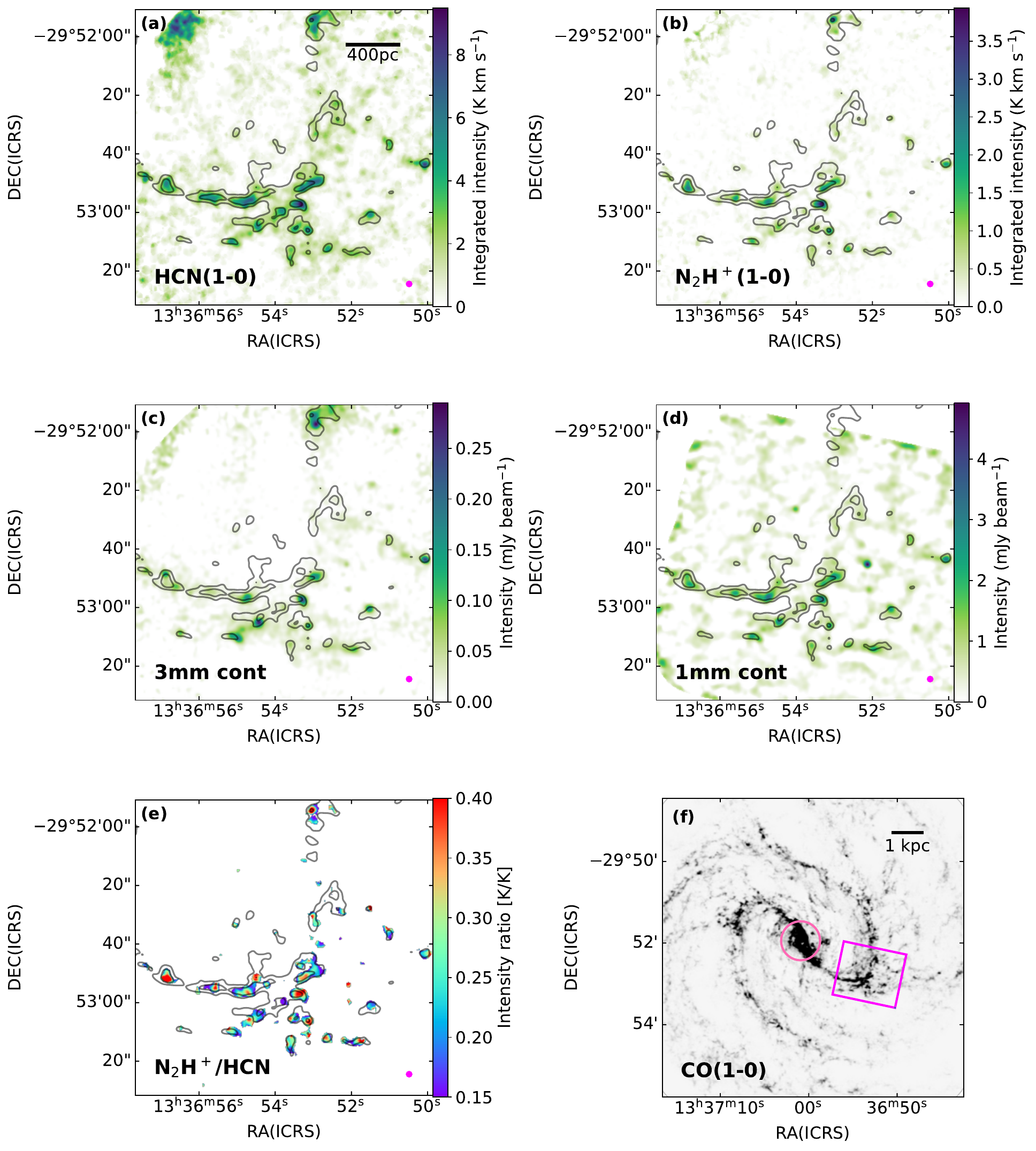}
\caption{The integrated-intensity images of HCN (a) and \nthp\ (b) and continuum images at wavelengths of 3\,mm (c) and 1\,mm (d). Panel (e) shows the ratio of the integrated intensities of \nthp\ to HCN. CO(1--0) intensity contours at 50 and 100\,K\,km\,s$^{-1}$ are overlaid, and the beam size is shown as magenta circles on all panels except for Panel (f). Panel (f) indicates the observed area marked by a magenta rectangle and the observing area of the archival data presented in Section \ref{sec:gal_cen} marked by a pink circle, overlaid on the CO moment 0 image from \citet{2023ApJ...949..108K}. We did not apply an rms cutoff to these maps on Panels (a--d). Maps of HCN and 3-mm continuum have already been presented previously in \citet{2024ApJ...969...82H}. \label{fig:mom0}}
\end{figure*}

Figure \ref{fig:mom0} presents velocity-integrated images of HCN (Panel a), \nthp\ (Panel b), continuum emission at wavelengths of 3\,mm and 1\,mm (Panels c \& d), and a ratio map of the integrated intensities of \nthp\ over HCN in the unit of K\,\kms (Panel e). We created the velocity-integrated images of HCN by applying a threshold mask. To construct this mask, we first convolved the original image to twice the beam size, and applied the threshold of 2$\sigma$. This mask was also applied to the \nthp\ velocity-integrated image because we expect regions with \nthp\ emission to also produce HCN emission. The typical number of channels integrated is 10, and typical rms values of integrated images are 0.3 K\,\kms\ for HCN and 0.1 K\,\kms\ for \nthp. The peak temperatures are the factors that contribute to the integrated intensities of CO, HCN, and \nthp\ because the velocity dispersions are relatively constant ($\Delta v(\mathrm{FWHM})=$12--17 \kms) in regions where a dense-gas tracer is detected. We note that the emission of HCN along the northeastern edge of the map is most likely an artifact, a noise component that happens to stretch over a wide frequency range. We excluded this part from scatter plots in subsequent sections.

The 3-mm continuum may contain multiple sources of emission such as free-free emission, dust emission, synchrotron radiation, and anomalous microwave emission (AME). The dominant source around $\lambda=3\,$mm is usually the free-free emission. The contamination from dust emission can be estimated using the 1-mm continuum image using the spectral index obtained from observations in the literature \citep{2012MNRAS.421.2917F}. We found that the contamination is mostly $\lesssim$10--15\,\%, but the fraction can go up to 30\,\% for very limited pixels. Meanwhile, none of the pixels have dust emission being dominant. Since there is no significant synchrotron source near the observed regions, and galaxies with AME features are rare, it is reasonable to only consider dust contamination. We also analyzed the observational trends shown in the next section after the contamination was removed, but the results only changed slightly. Because removing the contamination forces us to eliminate a part of the field of view not covered by the Band 7 observation, we keep the original 3-mm data for analysis.

The distribution of \nthp\ integrated intensities (Figure \ref{fig:mom0}b) appears to be more compact than that of HCN (Figure \ref{fig:mom0}a), although the peak signal-to-noise ratios are similar ($\sim 20$--$30$) for both transitions. The 3-mm and 1-mm continuum intensities (Figure \ref{fig:mom0}c and d) are also nearly as compact as the \nthp\ emission. The peaks of HCN, \nthp, and the 1-mm continuum coincide with the CO peaks, although CO peaks may not necessarily contain their peaks. 

In contrast to other emission, some of the 3-mm continuum peaks are slightly offset from the CO peaks by a few arcseconds, although others are co-spatial. It has been known from observations in the Local Group that locations of molecular gas and those of star formation are no longer co-spatial on scales less than 100\,pc \citep[i.e., the Kennicutt-Schmidt relation breaks down; ][]{2010ApJ...722L.127O,2019Natur.569..519K}. The offsets we see in our observations likely represent a similar trend.

For the \nthp/HCN ratio map (Figure \ref{fig:mom0}e), rms threshold values of $3\sigma$ and $5\sigma$ in integrated intensities are applied to \nthp\ and HCN, respectively. Visually, the ratio map appears to show a trend in which the ratios are higher in regions of the highest 1\,mm continuum intensity.

\subsection{Relationship among measured quantities}\label{sec:relation}
      
\begin{figure*}
\centering{
\includegraphics[width=0.8\textwidth]{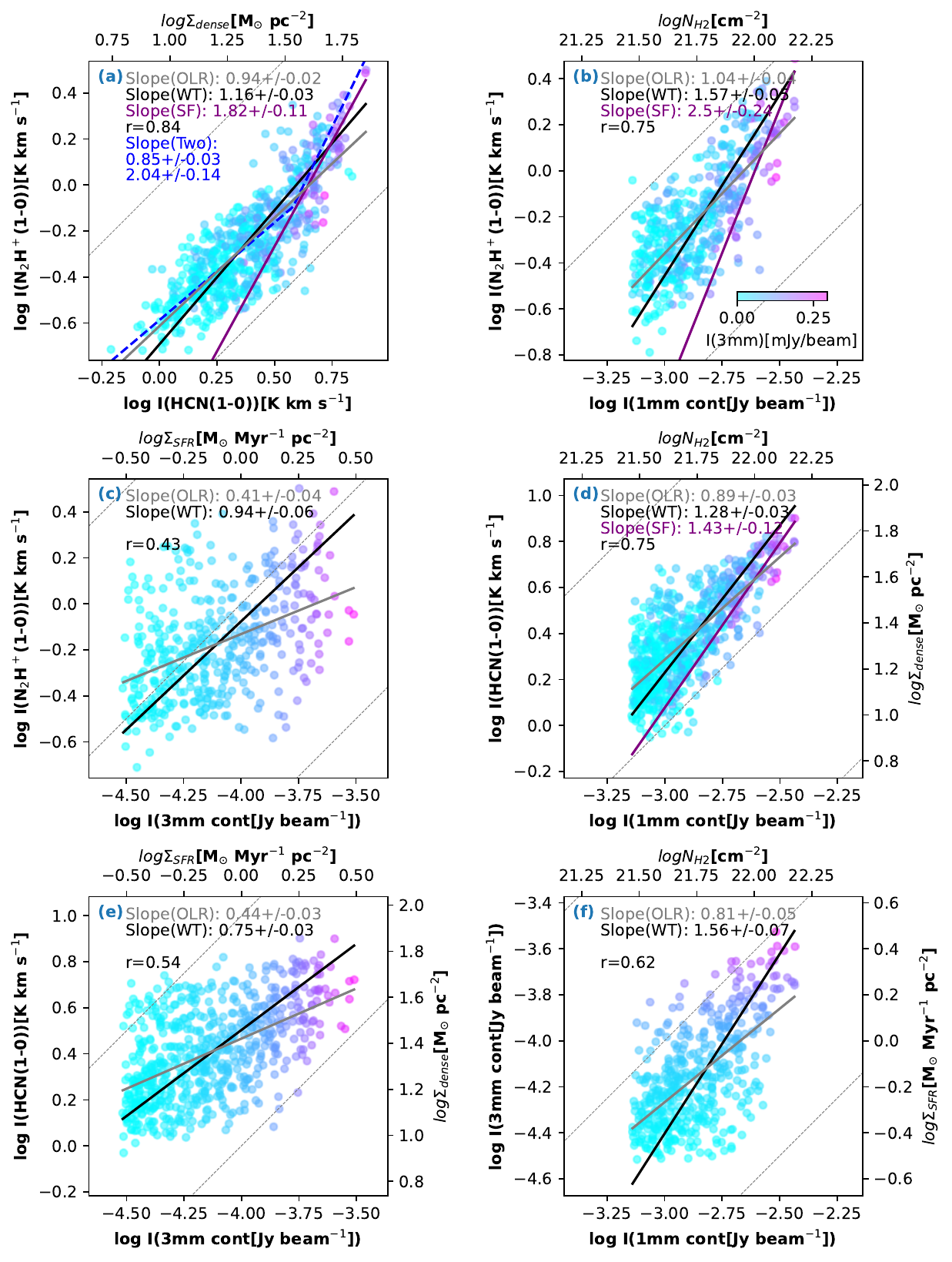}
}
\caption{Relationships between HCN and \nthp\ intensities, alongside 1-mm and 3-mm continuum emission. Secondary axes are shown when the primary axes are HCN, 1-mm, or 3-mm continuum, indicating the dense gas surface density, the column density, and the star formation rates, respectively. We explain the conversion methods in Appendix \ref{app:phys}. The black solid lines indicate the linear fits to the data points with signal-to-noise weighting (ODR), while red lines show fits without weights. Colors of the scatter points indicate the 3mm intensity. When the data points do not have $3\sigma$ detection of the 3-mm continuum, we zero-padded the 3-mm intensity values to show their colors when the 3-mm continuum is not used as an axis. The purple solid lines are the fit to data with \Itr$>0.15\,$mJy beam$^{-1} = 10^{-3.82}\,$\Jypbm\ and are only shown when the 3-mm continuum is not used for the plot axes. For Panel (a), we also show the fit for high HCN intensities as a black dashed line. \label{fig:scatter}}
\end{figure*}

Figure \ref{fig:scatter} shows the relationships among the integrated intensities of HCN, \nthp, and the intensities of the 1-mm and 3-mm continua. To avoid oversampling, we binned the data to a pixel size of $0\farcs9$ from the original size of $0\farcs3$ when examining these relationships. We applied the rms cutoff of 5$\sigma$ for HCN, and 3$\sigma$ for \nthp\ and 1-mm and 3-mm continua. In the 1-mm continuum data, there is an unresolved source that seems to be a higher-redshift sub-millimeter galaxy (SMG) around ($\alpha$,$\delta$)=($13^\mathrm{h}36^\mathrm{m}52^\mathrm{s}$,$-29^\circ 52\arcmin 40\arcsec$), which does not seem to have a counterpart in other maps. We exclude this SMG candidate from our analysis. The luminosities of HCN, the 1-mm continuum, and the 3-mm continuum can be used as proxies for the dense gas surface density, the molecular column density, and the star formation rate surface density, respectively. Using these tracers, we also calculate the corresponding physical quantities based on the equations provided in Appendix \ref{app:phys}, and display them on the secondary axes of Figure \ref{fig:scatter}. Please note that these physical parameters have some uncertainties. For example, the derived total column density depends on the assumed dust temperature and the missing flux, and a different $X_\mathrm{CO}$ conversion factor leads to a different molecular gas surface density. For each panel, we are showing two quantities used for axes, and the number of data points differs among panels.

We performed linear fitting of the log-log data points with the signal-to-noise ratio (S/N) weighted fit using the package \texttt{ODR}. Although the fit with weights should produce more reasonable results in most cases, we also refer to results using the \texttt{linregress} package from \texttt{scipy} to compare with the literature. Note that the slope of the fit in the log-log plane is the exponential, i.e., $log(y)=a\times1og(x)+b$ means $y=10^bx^a$ on the linear scale. We refer to the relation being super-linear when $a>1$, using the expression on the linear scale. Despite the small formal uncertainties on the derived slope for the correlations presented in Figure 2, the large scatter and low correlation coefficients indicate that the underlying relationships are often weak. We also note that continuum observations have clear detection limits. We examined their effect on the slopes by treating all the data points $2\sigma < I <3\sigma$ as upper limits with a cutoff of $3\sigma$ using a package \texttt{linmix} \citep{2007ApJ...665.1489K}. This test did not produce any significant difference from the ODR results.

The \nthp--HCN relation, meaning the relationship of \nthp\ as a function of HCN, (Figure \ref{fig:scatter}a), has the highest correlation coefficient among all the relationships discussed here, indicating that the relationship between these two dense gas tracers is tighter than the other correlations that we consider. Its slope is slightly super-linear (slope $\sim1.2$) for the S/N weighted fits, while it is almost linear without weights. As mentioned earlier, we applied a 5-$\sigma$ cutoff to HCN and a 3-$\sigma$ cutoff to other quantities. In addition to the rms ($\sigma$) of the measurement, a calibration error of 5\% is added when calculating the total error used to compute the S/N values. Visual inspection suggests that there appears to be a break in the power-law of this relationship, with a steeper slope at higher intensities. We examine this break by fitting the data with two different slopes connected at the break point. We chose the break point that minimizes the $\chi^2$ of the two-slope fit. For the lower HCN intensity part, the slope is slightly sub-linear, whereas it becomes super-linear with a slope of $\sim 2$ at the higher HCN intensity part. This behavior resembles the trend reported by \citet{2023A&A...680L..20S}, where the super-linear relation emerges at the higher end of HCN intensity. We further examined the case where we only considered pixels with high 3-mm luminosities ($I(\mathrm{3mm}) > 0.15$\,mJy\,beam$^{-1}$) to study the trend in active star-forming regions. The relationship is also super-linear for the active star-forming regions (slope $\sim1.8$).

This super-linearity can imply that the \nthp\ is tracing denser regions than HCN. Intensities of HCN and \nthp\ in general increase in regions with higher column densities. These regions also tend to have higher densities. If the \nthp/HCN ratios become higher in these regions, it is likely that \nthp\ is emitted more in denser regions than HCN. This point should also be applicable to other super-linear relationships discussed later.

Both HCN and \nthp\ intensities exhibit a super-linear relationship with the 1mm continuum for the weighted fit and at higher 3-mm luminosities, although they are both near-linear without weights (Figure \ref{fig:scatter} b \& d). This difference indicates steeper slope at higher 1mm continuum intensities. The super-linearity is more pronounced for \nthp\ than for HCN, having steeper slopes. The 3-mm continuum intensity shows weaker correlations with the other quantities (Pearson's correlation coefficients $r \sim 0.5$) compared to the stronger correlations among the other variables ($r \sim 0.8$). These weaker correlations are consistent with the spatial offsets observed between the 3-mm continuum and CO peaks in Section \ref{sec:images}, while \nthp, HCN, and the 1-mm continuum peak at similar locations to those of CO. A tighter correlation of the HCN intensity than the \nthp\ intensity with the 3mm continuum is consistent with Galactic observations, which show that HCN is associated with warm or UV-irradiated gas \citep{2020A&A...635A...4H,2023A&A...679A...4S} while \nthp\ emits more in cold gas \citep{2014A&A...568A..27F}. Still, the difference in correlation coefficients ($r=0.54$ vs $r=0.43$) is not large enough to consider it highly significant.

\begin{figure*}
\centering{
\includegraphics[width=0.8\textwidth]{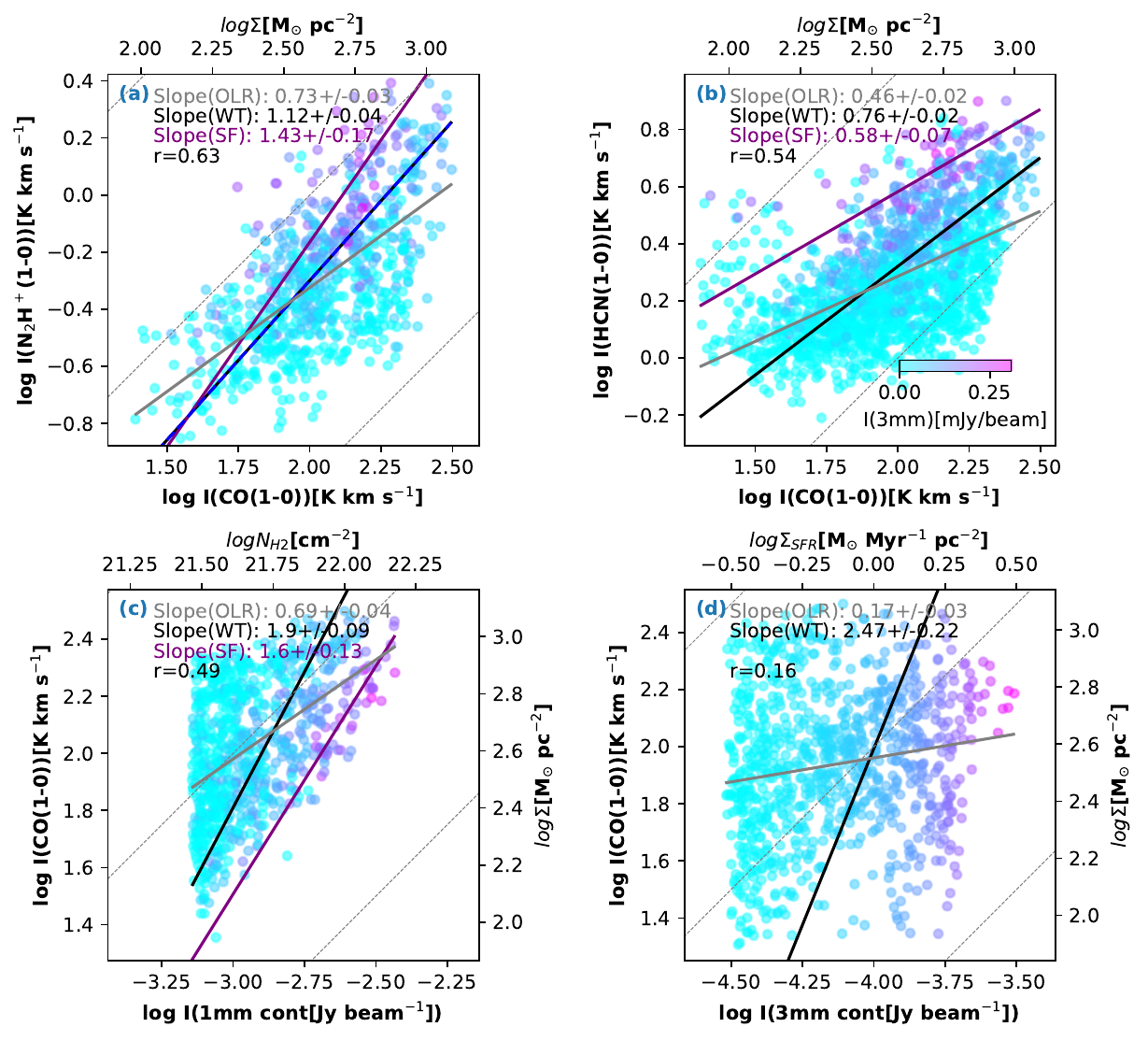}
}
\caption{Same as Figure \ref{fig:scatter}, but with CO intensities. Note that $1 M_{\odot}\,$pc$^{-2}$ corresponds to $N_{H2}=10^{19.7}\,$\sqc. \label{fig:scatter_co}}
\end{figure*}

The relationships between CO intensities and other quantities are shown in Figure \ref{fig:scatter_co}. In general, these correlations are weaker than those among the other quantities. For instance, the correlation between \nthp\ and CO is weaker ($r = 0.63$) than that between \nthp\ and HCN ($r = 0.84$) or between \nthp\ and \Ione\ ($r = 0.75$). This is likely because $^{12}$CO emission becomes easily optically thick and saturated even at moderately low column densities \citep[e.g., ][]{2023A&A...679A.112T}. This saturation is most directly demonstrated by the CO--\Ione\ relationship, particularly in the upper left portion of Figure \ref{fig:scatter_co}(c), mostly represented by regions with low $I$(3\,mm). In contrast, data points with high $I$(3\,mm) somehow exhibit a relatively tight correlation. In these regions with high star-forming rates, the dust temperatures may be higher, which overestimates the column density. These data points may represent lower column density, optically thinner regions. The data points in the bottom right part of Figure \ref{fig:scatter_co} (panels a \& b) also likely show this saturation regime. This region happens to have low 3-mm luminosity in general. The reason that regions with high star formation rates show fewer signs of saturation is that these regions have higher temperatures, and CO(1--0) may be optically thinner than cold regions. Because of the weak correlation, the S/N weighted fit between the CO intensity and the 3-mm continuum does not appear physical.

A similar explanation of CO saturation applies to the tighter correlation between HCN and the 1-mm continuum, compared to that between HCN and CO. In the HCN--CO relation, some data points lie in a regime with low CO and high HCN luminosities. These points are difficult to reconcile if the HCN strictly traces only dense gas. However, these points can be explained by the enhanced HCN luminosities in regions with high ultraviolet radiation field $G_0$ \citep{2017A&A...604A..74S} or low visual extinction $A_\mathrm{V}$ \citep{2023A&A...679A...4S}. The overall correlation between CO and \Itr\ is weak, likely due to the spatial offsets between their emission peaks.

\begin{figure*}
\centering{
\includegraphics[width=0.8\textwidth]{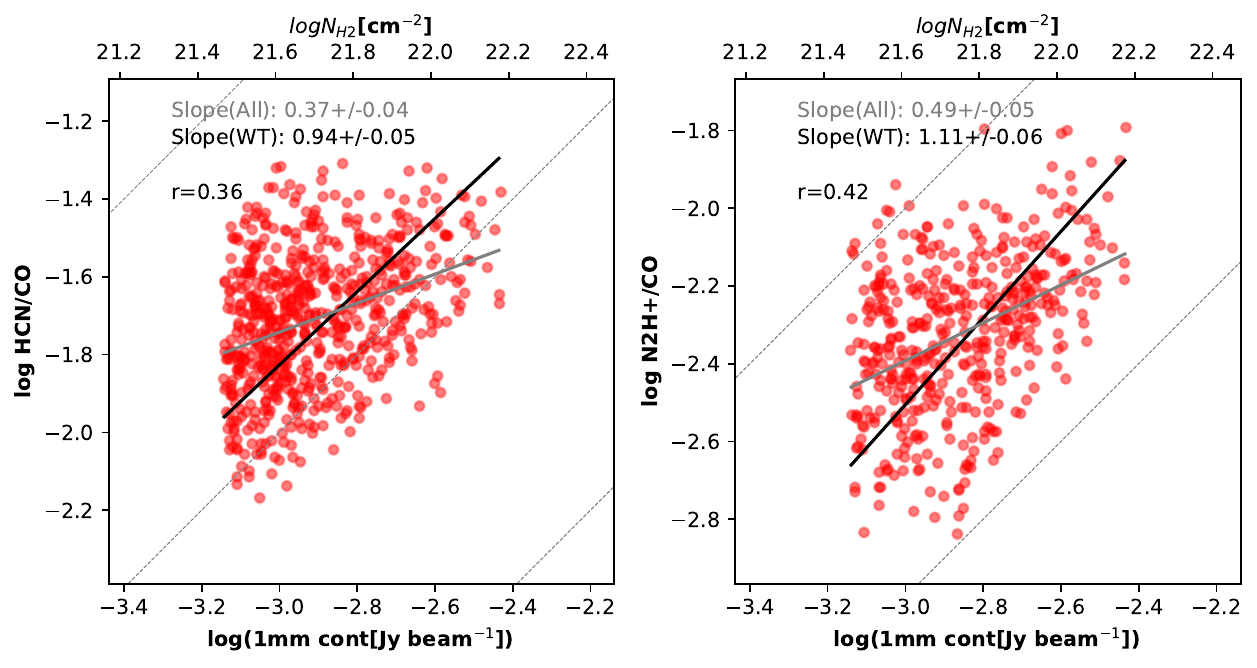}
}
\caption{The HCN/CO ratio (left) and the \nthp/CO ratio (right) as a function of the 1-mm continuum intensity. \label{fig:scatter_ratio}}
\end{figure*}

Figure \ref{fig:scatter_ratio} shows the HCN/CO ratio (left panel) and the \nthp/CO ratio (right panel) as a function of the 1-mm continuum intensity. Previous literature shows that the HCN/CO ratio has a positive correlation with the total column density. Galactic observations showed relatively little scatter with a slope of 0.7 \citep{2023A&A...679A.112T} using 0.03-0.06\,pc beam sizes. Meanwhile, extragalactic observations tend to have higher scatter \citep{2018ApJ...858...90G,2019ApJ...880..127J} with slopes of 0.4--0.8 using beam sizes of 300\,pc -- 2\,kpc. Larger scatter is most pronounced in the spatially resolved study by \citet{2025arXiv250719439S}, where the slopes are flatter or even negative depending on the location in M51. Similar to other extragalactic studies, our result also shows large scatter with the slope of $\sim 0.4$. The S/N weighted fit has a higher slope, but we have to be cautious before taking it at face value. This is because the S/N weighted fit does not always provide a more reasonable fit when the scatter is large and high S/N data points influence the fit too strongly. The \nthp/CO versus the 1-mm continuum intensity plot also shows a positive correlation with a slightly steeper slope.

One may wonder if it is helpful to compare emission intensities discussed above with the CO(2--1)/(1--0) ratio ($=R_{21}$) in \citet{2025arXiv250508876K}. We note that it is difficult to obtain meaningful information on densities or temperatures from such a comparison because CO emission is usually optically thick in regions with \nthp\ and HCN detections. The $R_{21}$ values are relatively constant at high values (=0.7--0.9) in regions with dense gas detections, which is expected from optically thick CO emission.

\section{Radiative transfer modeling}\label{sec:rad}

If \nthp\ is more efficiently emitted in dense regions than HCN, the power-law index of the \nthp--HCN relationship should be super-linear. The trend we observed in Section \ref{sec:relation} between \nthp\ and HCN, as well as between \nthp\ and the 1-mm continuum, is consistent with the idea that \nthp\ intensity is enhanced in denser regions. However, we still lack a quantitative understanding of the observed trend. To investigate this relationship quantitatively, we used the publicly available radiative transfer code RADEX \citep{2007A&A...468..627V} to compute column density--dependent line intensities.

\subsection{Model setup}
Our model setup has two components. The first step is the computation of transition intensities for each column density from the molecular abundances as functions of column density. The second step is to add up the results of the first step according to certain probability distribution functions (PDFs) of column densities.

We used the column-density--dependent abundance model introduced by \citet{2021A&A...646A..97T}, as described in Appendix \ref{app:chem}, based on their observational results in the Perseus Molecular Cloud.  We used the volume density as a function of column densities also from \citet{2021A&A...646A..97T} shown in Equation \ref{eq:dens}. We use the temperature of 20\,K for CO and HCN, while we use 10\,K for \nthp. In high-mass star-forming regions, the gas temperature can be warmer, so we use 20\,K for CO and HCN. The colder temperature of \nthp\ is justified by the fact that this molecule emits from colder regions \citep[e.g., ][]{2014A&A...568A..27F}. 

We used different line widths (FWHM) for each line for the RADEX \citep{2007A&A...468..627V} modeling. The observed line width is $\sim10-20\,$km s$^{-1}$ in our observations, but the large dispersion may be due to large-scale motions across the molecular clouds. For the case of HCN and \nthp, hyperfine structure lines may also contribute to larger observed line widths. The effective line widths for the radiative transfer at each column density bin are likely to be narrower. Line widths of HCN and \nthp\ can be as narrow as 1--2\,\kms\ in observations of Galactic molecular clouds. We use the collisional coefficients from the LAMDA database \citep{2005A&A...432..369S} without the distinction of hyperfine components because it is not correct to simply sum up the intensities of each hyperfine line, which may overlap with each other. However, we are also overestimating the optical depth if we do not make a distinction between hyperfine components. Therefore, we use the line widths of 5\,\kms\ for HCN and \nthp\ to account for the hyperfine structure lines. For CO, we use the line width of 10\,\kms\ because its line width in Galactic observations tends to be a factor of several higher than other transitions because it is optically thick.

\begin{figure}
\centering{
\includegraphics[width=0.47\textwidth]{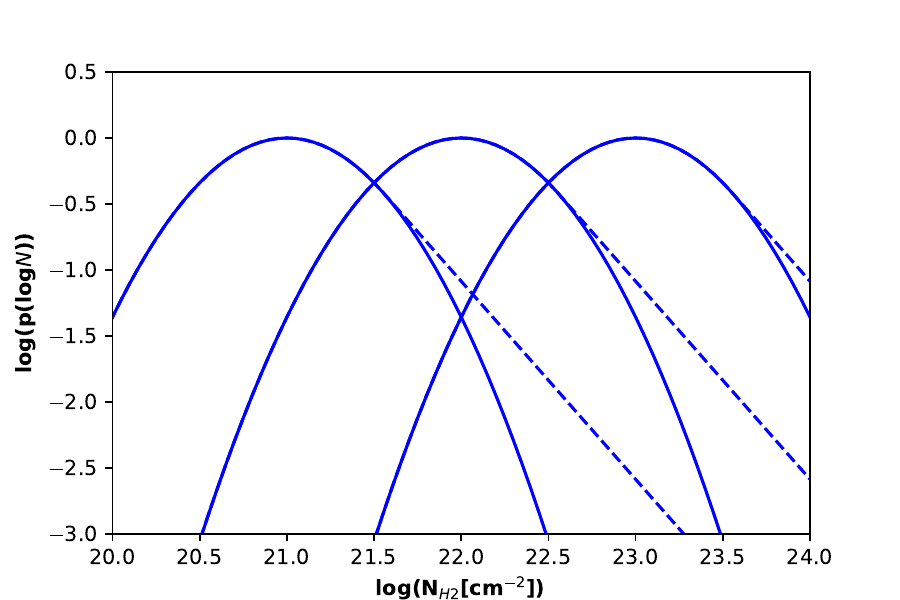}
}
\caption{Column density distributions used for the radiative transfer model for the log-normal distribution together with the power-law extension at the higher density end (dashed) and log-normal distribution only (solid) for log($N_\textrm{mode}$/\sqc)=21, 22, and 23, normalized by its peak value.\label{fig:column_model}}
\end{figure}

In extragalactic sources, variations in column density within GMCs cannot be resolved, as the beam sizes are typically comparable to the size of GMCs. To account for this, we considered the PDF of the column densities, adopting both a log-normal distribution and a log-normal distribution with a power-law tail at the high-density end (Figure \ref{fig:column_model}), motivated by observations of Galactic molecular clouds \citep{2009A&A...508L..35K,2013ApJ...766L..17S,2017ApJ...834L...1B}.

When the H$_2$ column density $N$ follows a log-normal distribution, the PDF of $x = \mathrm{ln}(N/N_0)$ is commonly expressed as
\begin{equation}
p(x) = \frac{1}{\sqrt{2\pi \sigma_{\eta,e}^2}} \mathrm{exp}\left[-\frac{1}{2}\left(\frac{x-\mu}{\sigma_{\eta,e}}\right)^2\right]    
\end{equation}
where in $x = \mathrm{ln}(N/N_0)$, $N$ is the column density under the log-normal distribution, $N_0$ is the mean (expectation value) of the column density, which requires that the expectation value of $x$ to be $\mu = -\frac{1}{2}\sigma_{\eta,e}^2$. The standard deviation of $x$ is denoted as $\sigma_{\eta,e}$. The subscript on $\sigma$ indicates that it pertains to the column density rather than the volume density, and that it is expressed in natural logarithms. The mode (most frequent value) of the PDF of the logarithm of $N$ is at $\mathrm{ln}N_{\rm mode} = \mathrm{ln}N_0 +\mu= \mathrm{ln}N_0 - \frac{1}{2} \sigma_{\eta, e}^2 < \mathrm{ln}N_0$, while the mode of the PDF of $N$ in the linear scale is at $N_0 e^{-\frac{3}{2}\sigma_{\eta,e}^2}$. We use the notation in the logarithmic scale throughout this paper.

Because it is more convenient to work in base-10 logarithms, we rewrite the PDF as:
\begin{equation}
 p(y) \propto  \mathrm{exp}\left[-\frac{1}{2}\left(\frac{y - y_0 - \mu'}{\sigma_{\eta,10}}\right)^2\right]    
\end{equation}
where $y = \mathrm{log}N$, $y_0 = \mathrm{log}N_0$, $\sigma_{\eta,10} = \frac{1}{\mathrm{ln}10}\sigma_{\eta,e}$, and $\mu' =\frac{1}{\mathrm{ln}10}\mu= -\frac{1}{2}\,\mathrm{1n10}\,\sigma_{\eta,10}^2$ as $\mu=-\frac{1}{2}\sigma_{\eta,e}^2$. For simplicity, we also define $y_{\mathrm{mode}} = (\mathrm{log}N)_{\mathrm{mode}} = y_0 + \mu'$, representing the peak of the PDF. The PDF again peaks at $y_{\rm mode} = y_0 + \mu^\prime = (\log N)_{\rm mode}$. We note that the value of $\mu'$ is small ($\lesssim 0.2$ for a reasonable $\sigma_{\eta,10}$), and thus the difference between $y_0$ and $y_{\mathrm{mode}}$ is also small.

The power-law tail of the PDF is expressed as
\begin{equation}
    p(y)\propto \mathrm{exp}(\alpha \mathrm{ln}N) = N^{\alpha} = 10^{y\alpha},
\end{equation}
where $\alpha$ is the slope of the PDF in a logarithmic column density plot. We switch from the log-normal to the power-law distribution where the derivatives $dp(y)/dy$ of the two functions become equal, for $N > N_{\mathrm{mode}}$. As fiducial values, we adopt $\sigma_{\eta,10} = 0.4$ and $\alpha = -1.5$. The rationale behind these parameter choices is discussed in Appendix \ref{app:logn}.

We use the column-density-dependent intensities described above for each $N$ (or $y$), $I(y)$, and integrate them over the column-density PDF:
\begin{equation}
    \overline{I}(\sigma_{\eta,10}, y_{\mathrm{mode}}) = \frac{\int p(y, \sigma_{\eta,10}, y_{\mathrm{mode}}) I(y) \, dy}{\int p(y, \sigma_{\eta,10}, y_{\mathrm{mode}}) \, dy}.
\end{equation}

We ran RADEX for the range of log$N/\mathrm{cm}^{-2}=20-24$, and found that the emission is insignificant in contributing to the overall emission at log$N/\mathrm{cm}^{-2}=20$, justifying the minimum column density of computation. The maximum column density of computation at log$N/\mathrm{cm}^{-2}=24$ is determined due to the lack of Galactic observations beyond this column density, and it is expected that the extrapolation of fractional abundances will fail in this regime. Subsequently, we constructed the column density distributions for log($N/\mathrm{cm}^{-2}$)$_{\mathrm{mode}}$=20.5--23. 
The mean (i.e., expectation value) of the column density, $\langle N_{H2} \rangle$, differs from $N_0$ under the log-normal + power-law distribution, and we integrated the column-density-weighted PDF over log$ (N/\mathrm{cm}^{-2}) = 18.5-24$ to calculate its value.

This approach is conceptually similar to the method employed by \citet{2017ApJ...835..217L}, although they assumed constant fractional abundances. \citet{2025A&A...703A.176Z} recently developed a method to constrain the column density PDF parameters based on the line intensities observed towards the Orion B region using the Bayesian inversion algorithm, and applied it to a part of the observed area in M51. Unlike \citet{2025A&A...703A.176Z}, our aim is not to constrain the parameters but to examine the variation with parameters. A detailed exploration of a similar methodology will be presented in Nishimura et al. (in preparation).

Note that this PDF treatment also reproduces the difference in the beam filling factor among different molecular transitions. Within the 50-pc beam, the beam filling factor of CO should be the largest, and the ones for HCN and \nthp\ are expected to be smaller. A single PDF contains components with enough column densities or volume densities to emit all of \nthp, HCN, and CO, components that emit CO and HCN, ones that emit only CO, and ones that do not even emit CO. These different components naturally create different beam filling factors, which should be present if we were to observe them with higher angular resolutions.

\subsection{Modeled intensities}

\begin{figure*}
\centering{
\includegraphics[width=0.9\textwidth]{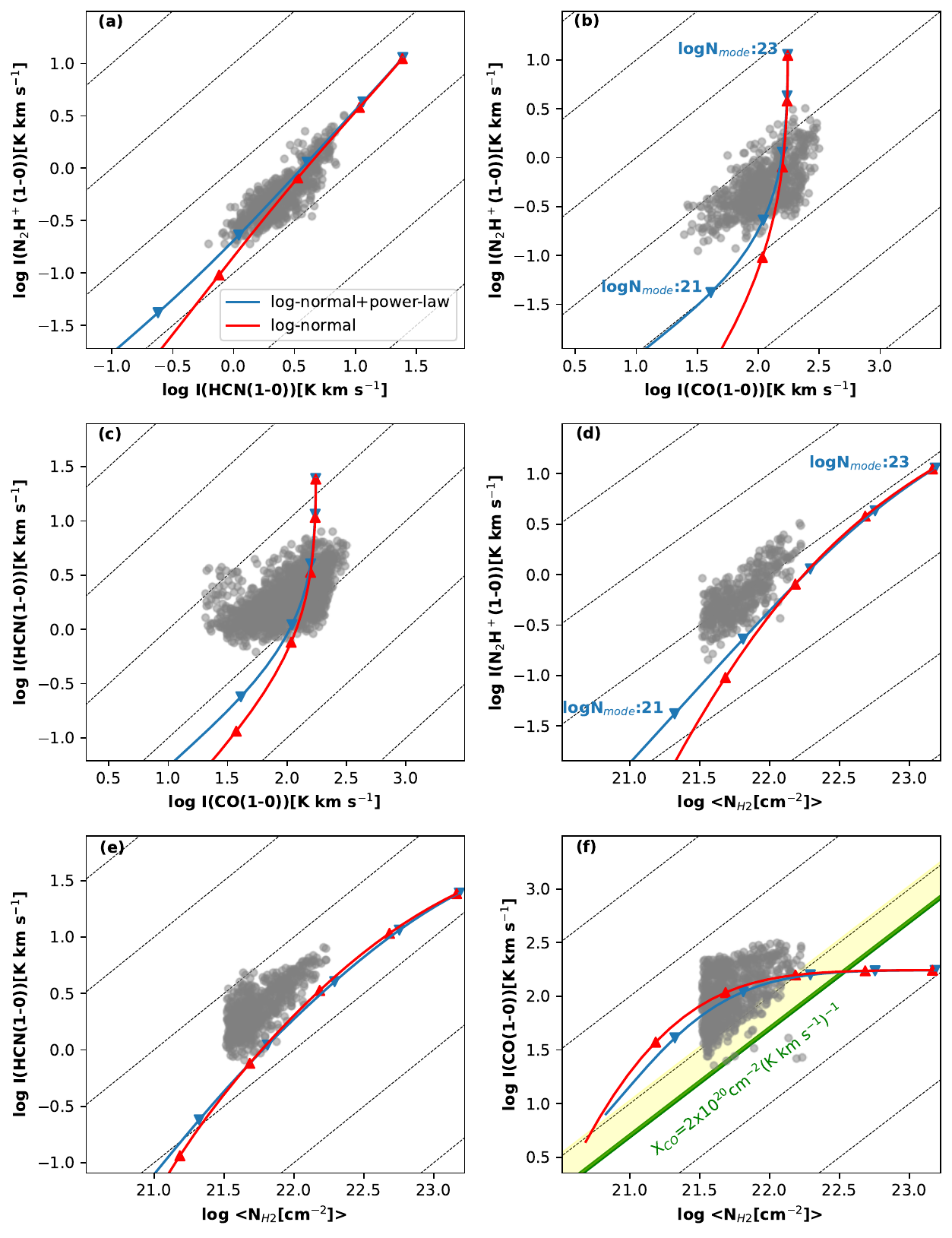}
}
\caption{Modeled intensities for log-normal+power-law (blue) and log-normal (red) column density distributions overlaid on the observed data points. Markers are plotted for log$N_\textrm{mode}$=[20.5, 21.0, 21.5, 22.0, 22.5, 23.0] with some values labeled in the top right panel. We also plot the black dashed lines with slopes of 1 with 1 dex. increments to guide the eye. The green line in panel (f) shows a typical Milky Way $X_{CO}$ conversion factor \citep[$2\times 10^{20}\,$cm$^{-2}$(K\,\kms)$^{-1}$;][]{2013ARA&A..51..207B}, and the yellow area shows the conversion factor derived by \citet{2024ApJ...968...97L} for $R=3$--4\,kpc. \label{fig:model}}
\end{figure*}

Figure \ref{fig:model} shows modeled $\overline{I}(N_\mathrm{mode})$ as the peak column density is varied over the range $\log(N/\mathrm{cm}^{-2})_{\mathrm{mode}} = 20.5$--$23$, for both log-normal-only and log-normal plus power-law ($\alpha = -1.5$) distributions. In the figure, observational data are also shown. In panels (d)-(f), the column densities in the abscissa are $\langle N_\mathrm{H_2}\rangle$ for the models and those estimated from the 1mm continuum using Equation \ref{eq:hildebrand} in the Appendix for the observational data.

In the \nthp--HCN plot (Figure \ref{fig:model}a), both the log-normal-only and log-normal+power-law models predict the observed relationship well. The log-normal-only model has a slightly super-linear slope ($\sim1.4$), while the log-normal+power-law model is almost linear (slope $\sim 1.2$). The modeled relationships between molecular line intensities and total column densities seem to deviate from the observed ones (Figure \ref{fig:model} d-f). The deviation suggests that either the model underestimates the molecular line intensities or the observation underestimates the total column density from the dust observation. The model can underestimate line intensities if molecular transitions emit differently for a given H$_2$ column density in M83 from the Galactic template used for the model. The observed column densities can be underestimated if we assume a higher dust temperature than the actual value. Dense regions emitting \nthp\ may have lower dust temperatures than the one derived from the lower-resolution Herschel observation the dust observation \citep{2012MNRAS.421.2917F}. In addition, the dust opacity has the uncertainty of a factor of 2--3 (see Appendix \ref{app:phys}), which can linearly affect the estimation of the total column density.

The modeled relationships involving CO (i.e., \nthp-CO, HCN-CO, and CO-$\langle N_{\mathrm{H}_2} \rangle$; panels (b), (c), and (f), respectively) show that CO emission is saturated in the optically thick regime (log($N_{peak}/\textrm{cm}^{-2})\gtrsim \,21.5$). Meanwhile, the observed relationships between CO and other quantities exhibit positive correlations. This difference between models and observations of relationship of CO and other quantities may arise because higher temperatures or increased velocity dispersion in high column density regions enhance CO intensities linearly in optically-thick regions as discussed in Appendix \ref{app:plots}. The velocity-integrated intensity of CO increases with a larger velocity width because CO is usually optically thick at each velocity. Because the observed velocity dispersion is relatively uniform across regions with HCN or \nthp\ detection, the variation in CO intensity at high column densities is likely due to elevated temperatures. However, the line-of-sight velocity dispersion may not be the intrinsic or local velocity dispersion, which may have uncertainties of a factor of a few. Because the temperature can also vary in molecular clouds by up to a factor of 5 \citep{2017A&A...599A..98P,2023A&A...679A.112T}, the discrepancies between the models and the observations of factors of a few are still within the expected range.

In the \nthp--$\langle N_{\mathrm{H}_2} \rangle$ relationship, both models predict the super-linear relationship as seen in the observations. However, the \nthp\ emission in the model is underestimated for the given column density. The models also predict the slope of the HCN--$\langle N_{\mathrm{H}_2} \rangle$ relationship well, while there is an offset of a factor of $\sim2$ ($\sim0.3$ dex) between the observed points and models similar to the \nthp--$\langle N_{\mathrm{H}_2} \rangle$ relationship. This discrepancy between models and observations in relations between the N$_2$H$^+$ or HCN intensities and the column densities may result from the uncertainty in the derivation of the column density from the dust observations as discussed earlier.

\begin{figure*}
\centering{
\includegraphics[width=0.9\textwidth]{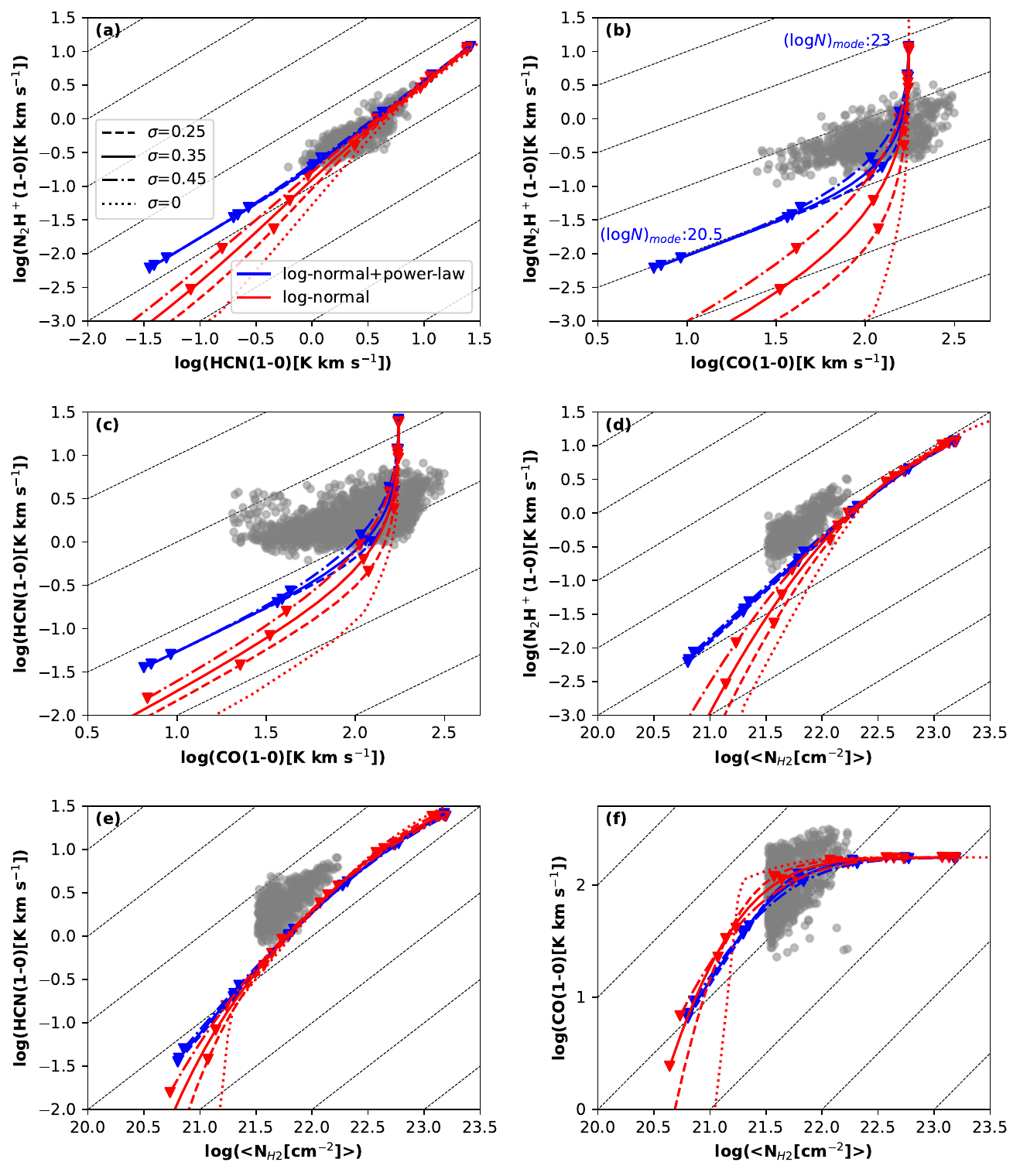}
}
\caption{Modeled intensities with varying $\sigma_{\eta,10}$ from 0.25 to 0.45 with the power-law of $\alpha=-1.5$ (blue) and without power-laws (red). Black dashed lines with slopes of 1 are plotted to guide the eye. \label{fig:model_sigma}}
\end{figure*}

In addition to the fiducial model above, we also evaluate how the model PDF parameters influence the line intensities. We also show wider ranges than our observed values in figures so that we can make predictions. First, we varied $\sigma_{\eta,10}$ within the range 0.25--0.45 (Figure \ref{fig:model_sigma}). This range is reasonable for Mach numbers $\mathscr{M} = 5$--$100$, based on the rough estimate described in Appendix \ref{app:logn}. We also show intensities of single column densities without considering the PDF, corresponding to the case of $\sigma_{\eta,10}=0$. As $\sigma_{\eta,10}$ increases, the predicted intensities also tend to increase for the same $N_\textrm{mode}$ and $\langle N_{H_2}\rangle$, particularly at lower values of $N_{\mathrm{mode}}$ for the log-normal model. Meanwhile, the change is insignificant for the model with the power-law tail. The case when $\sigma_{\eta,10}=0$ suggests that the single column density model underproduces \nthp\ and HCN with respect to CO or column densities.

We note that, in reality, an increased Mach number enhances both the dispersion of the column density PDF ($\sigma_{\eta,10}$) and the observed velocity dispersion ($\Delta v$). However, in this analysis, we vary only one parameter at a time to isolate its effect. If $\Delta v$ were varied instead, an integrated intensity of CO would increase nearly linearly with $\Delta v$ when CO is optically thick, while the HCN and \nthp\ integrated intensities would remain largely unchanged.

\begin{figure*}
\centering{
\includegraphics[width=0.9\textwidth]{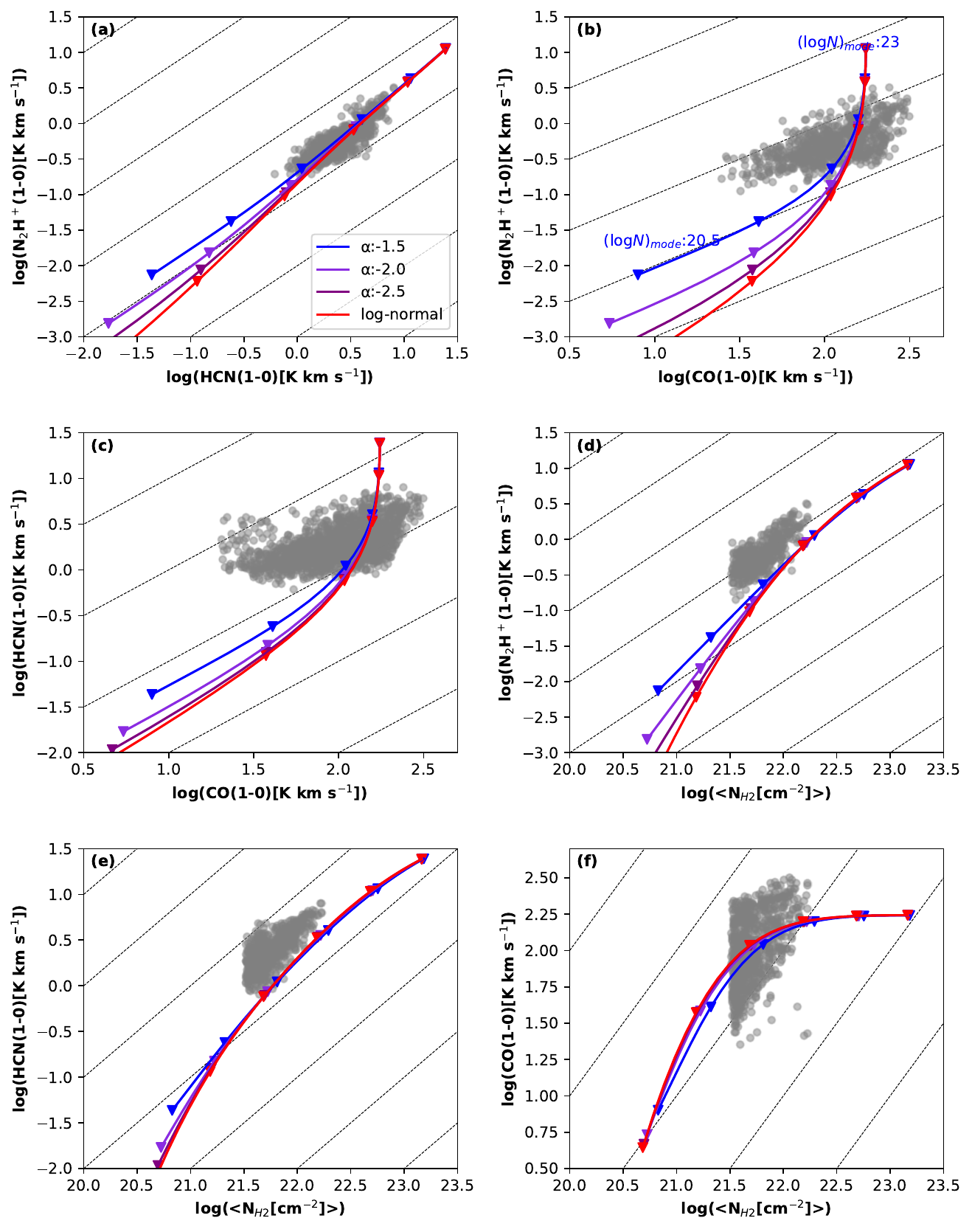}
}
\caption{Modeled intensities with varying $\alpha$ from $-2.5$ to $-1.5$ and with log-normal only when $\sigma_{\eta,10}=0.4$.  Black dashed lines with slopes of 1 are plotted to guide the eye. \label{fig:model_slope}}
\end{figure*}

We also examined the effect of varying the power-law slope $\alpha$ in the range from $-2.5$ to $-1.5$ (Figure \ref{fig:model_slope}). Among all transitions, the \nthp(1--0) intensity is most sensitive to changes in $\alpha$, as expected, since \nthp\ primarily traces the densest gas components. Although this result implies that we can use \nthp\ emission to detect the power-law tail from observations, the difference between the log-normal-only model and log-normal+power-law model is small when the average column density $\langle N_{H_2} \rangle \gtrsim 10^{21.5}\,$\sqc. This column density is below the detection limit in our 50-pc resolution observation.

In addition, plots in Appendix \ref{app:plots} show variations with the kinetic temperature, line width, and fractional abundances. The results are exactly what one would expect for optically thick/thin transitions. An optically thick line, such as CO, increases its intensity with increasing temperature or line width, while its increase is less for less optically thick lines. On the other hand, transition intensities linearly correlate with fractional abundances in optically thin transitions or regimes while the variation is less for optically thick cases.

\section{Discussion}\label{sec:disc}
\subsection{Column density distribution and observed relationships}\label{sec:dis_rel}
As presented in Section \ref{sec:results}, the \nthp--HCN relationship (Figure \ref{fig:scatter}a) is super-linear using the S/N weighted fit. The super-linearity becomes more pronounced when we fit them with two slopes. This is consistent with previous findings in M51, where the overall fit without weights shows a linear trend, while the fit among data points with high HCN luminosities becomes super-linear \citep{2023A&A...680L..20S}. The radiative transfer models using both log-normal and log-normal+power-law PDFs described in Section \ref{sec:rad} can reproduce slightly super-linear trends that match the observations well. This result suggests that molecular clouds in M83 can be represented by combinations of Perseus-like clouds with different mean column densities.

Meanwhile, our individual models did not reproduce the break in the power law. Observed data have steeper slopes at higher HCN luminosity, while our models have only a slight change in slopes, and this subtle difference likely results from variations among different molecular clouds. However, if we combine models with low and high $\sigma$ with and without the power law, it should be possible to generate the break in the power law. Furthermore, in intensity modeling, we adopted the fitting results from \citet{2021A&A...646A..97T}, where molecular line intensities are expressed as functions of column density. However, in reality, additional environmental factors can significantly influence molecular line intensities. While Perseus is a low-mass star-forming region, our field of view toward M83 likely also contains high-mass star-forming regions. The high-mass star-forming region Orion B has about a factor of a few higher HCN intensities than Perseus does \citep{2025A&A...703A.176Z}. Orion B also emits \nthp\ more preferentially from higher column-density regions (i.e., lower luminosity than Perseus at lower column-density regions, higher luminosity at higher column-density regions). This difference will shift the model to the right side of the plot in Figure \ref{fig:model}(a) with more super-linearity. It seems likely that we can explain the spread of the observed data points in M83 easily with the mixture of low- and high-mass star-forming regions.

 There are some possible causes of differences between low-mass and high-mass star-forming regions. High-mass star-forming regions tend to have a stronger UV radiation field, and possibly higher cosmic-ray ionization rates. Kinetic temperatures are also higher in high-mass star-forming regions. A stronger UV radiation field can enhance HCN emission \citep{2017A&A...604A..74S,2023A&A...679A...4S}, while elevated cosmic-ray ionization rates can increase the abundance of \nthp. In contrast, higher dust temperatures can suppress \nthp\ abundances by promoting the sublimation of CO, which subsequently destroys \nthp. Higher temperatures in Orion B may also lead to elevated \twco\ intensities. 

 The shape of the PDF can also change the intensity relationship. We employed the log-normal or a combination of log-normal and power-law PDFs. If the actual PDFs significantly differ from these ones, the \nthp-HCN relationship will have different slopes. For example, if the beam contains regions that are almost void of gas (i.e., a beam filling factor of less than unity), the PDFs will have double peaks. In this case, the points in Figure \ref{fig:model} shift downwards and leftward along lines with slopes of 1.

With the increasing availability of multi-line observations of molecular clouds on cloud scales \citep[Orion B, Orion A, W51, W3(OH), W49, Perseus, California;][]{2017A&A...599A..98P,2017A&A...605L...5K,2017ApJ...845..116W,2017ApJ...848...17N,2020MNRAS.497.1972B,2021A&A...646A..97T,2023A&A...679A.112T}, it is now crucial to compare line intensities as functions of column density, gas temperature, interstellar radiation field strength, kinetic temperature, and other measurable parameters. Robust modeling and reliable derivation of the column density PDF will only be possible after acquiring well-constrained intensity--column density relationships, along with their associated uncertainties, from such regions.

\subsection{Star formation and dense gas}
Our results in Figure \ref{fig:scatter}(a) show that the cold dense gas traced by \nthp\ has a steeper relationship with HCN and 1-mm continuum intensity in locations with high 3-mm continuum intensity than in other locations. Our model (Figure \ref{fig:model}a) suggests that column density distributions with the power-law tails have flatter relationships compared with the log-normal-only distributions. If the power-law tails lead to star formation, it appears that locations with high 3-mm intensities should have flatter relations, which is in contradiction with the observed trends.

We suggest that this trend may be due to the time evolution. Before star formation, the dense gas fraction and the column density both increase (i.e., data points move towards the top right direction in the \nthp-HCN and \nthp-1-mm plots; Figure \ref{fig:model}a and d). For some time after stars form, \nthp\ can co-exist with star formation (data points stay at the top right corner in the  \nthp-HCN and \nthp-1-mm plots), but the star-forming regions become devoid of cold gas after a while (data points move downward in the plots). 

If the above argument of timescale is valid, the cold dense gas must be spending some time without star formation, with both star formation and cold dense gas coexisting, and star formation with less cold dense gas. In other words, there should be comparable fractions of regions with \nthp-only, both \nthp\ and 3-mm continuum, and 3-mm continuum only detection. \citet{2019ApJ...887...49S} estimated the lifetime of molecular clouds by 
\begin{equation}
t_\mathrm{gas} = t_\mathrm{H\alpha} \frac{f_\mathrm{CO-only}+ f_\mathrm{overlap}}{f_\mathrm{H\alpha-only}+f_\mathrm{overlap}}=t_\mathrm{H\alpha} f_\mathrm{scale}
\end{equation}
where $t_\mathrm{gas}$ is the lifetime of molecular clouds, $t_\mathrm{H\alpha}$ is the timescale where HII regions are visible, $f_\mathrm{CO-only}$ is the fraction of pixels with CO detection without H$\alpha$, $f_\mathrm{overlap}$ is the fraction of pixels where both CO and H$\alpha$ are detected, and $f_\mathrm{H\alpha-only}$ is the fraction with H$\alpha$ detection without CO detection, while $f_\mathrm{scale}$ is introduced to simplify the expression. \citet{2023ApJ...959....1K} claim that this is not necessarily the lifetime of molecular clouds, and this timescale can be considered as the duty cycle of star formation. In either case, examination of the size of the area can give us some hint on the respective timescale. 

We can apply the same concept using CO, HCN, and \nthp to investigate the timescales of dense gas and star formation. We note that the only 3-mm continuum is available for us, not H$\alpha$, and we are likely only detecting the most active star formation, which results in a smaller area of star formation than there actually is. For CO, HCN, and \nthp, the sensitivity and applied cutoff should trace similar molecular clouds as we explain below. For \nthp\ and HCN, we used the same integrated intensity maps as those used for Figures \ref{fig:scatter} and \ref{fig:scatter_co} with sensitivity cutoffs applied. As mentioned previously, the typical rms values for HCN and \nthp\ are 0.3 K\,\kms\ and 0.1\,K\,\kms, respectively. With 5- and 3-$\sigma$\ masks, the cutoff levels are 1.5 K\,\kms\ for HCN and 0.3\,K\,\kms\ for \nthp. Because the intensity ratios $I$(HCN):$I$(\nthp)$\sim$5:1 from Figure \ref{fig:scatter}, these cutoffs correspond to similar molecular content. Similarly, because $I$(CO):$I$(\nthp)$\sim$200:1 from Figure \ref{fig:scatter_co}, the corresponding cutoff for CO should be 20\,K\,\kms. We use this cutoff value for the CO map when we evaluate the timescales, but the results do not change by using a lower cutoff value. For the case of CO, $f_\mathrm{CO-only}$ : $f_\mathrm{overlap}$ : $f_\mathrm{3mm-only}$=90\% : 9.9\% : 0.067\%, leading to $f_\mathrm{scale, CO}=10$. These values for HCN are $f_\mathrm{HCN-only}$ : $f_\mathrm{overlap}$ : $f_\mathrm{3mm-only}$=52\% : 33\% : 15\% with $f_\mathrm{scale, HCN}=1.8$, while they are $f_\mathrm{N_2H^+-only}$ : $f_\mathrm{overlap}$ : $f_\mathrm{3mm-only}$=27\% : 34\% : 39\% with $f_\mathrm{scale, N_2H^+}=0.84$. 

In addition to the pixel-based analysis above, we also tried a flux-based analysis. The pixel-based method works when the detection limit is low enough, and sources are resolved (e.g., CO and H$\alpha$). However, if the emission region is smaller than the beam size (i.e., the beam filling factor is less than unity), counting only the pixel number in high-flux regions underestimates the emission region there. If we calculate the fraction of a molecular intensity with the 3mm detection over that of the total intensity, we can estimate the fraction of time that the molecule spends hosting ongoing star formation. These fractions are 18\% for CO, 52\% for HCN, and 71\% for \nthp. 

These values imply that the timescale is shortest for \nthp\, and slightly longer for HCN, while it is far longer for CO. This result suggests that dense gas is more closely related to star formation. The formation of dense gas, the formation of stars, and the destruction of dense gas are likely taking place on comparable timescales. 

\subsection{Comparisons with other works}

There are only a handful of observations that have detected \nthp\ in the disk regions of galaxies. A previous study of M51 by \citet{2023A&A...680L..20S}, using a resolution of 125\,pc, found the slope of the \nthp--HCN relationship to be nearly linear at $\sim1.1$ ($\sim1.2$ excluding the AGN) across all data points, and steeper at $\sim1.5$ when fitting only the highest intensity regions. These findings are consistent with our results, which show an overall slope slightly super-linear($1.2$), and a higher slope ($1.5$) for regions with high 3-mm intensity or a two-slope fit ($2.0$), based on our 48-pc resolution data.

Our derived \nthp/HCN ratio of $0.25 \pm 0.08$ with 48-pc resolution is in agreement with the value reported for the disk of M51 with the 125-pc resolution, $0.24 \pm 0.26$, although our maximum values are higher. Approximately 1\% of our data points exhibit \nthp/HCN ratios in the range 0.5--0.72, whereas all data points in M51 remain below 0.5. These ratios are also higher than those found in kpc-scale observations of NGC~6946 by \citet{2023A&A...676L..11J}, who reported a ratio of $0.15 \pm 0.02$ using the IRAM 30-m telescope. A similar result was found by \citet{2014ApJ...788....4W} in their IRAM 30-m observations towards the disk region of M51, reporting \nthp/HCN ratios of $0.19 \pm 0.02$ and $0.14 \pm 0.05$ within a 1.25\,kpc beam for two different positions. These results indicate that higher-resolution interferometric observations tend to obtain higher \nthp/HCN ratios compared with single-dish observations. This trend is reasonable as the smaller beam sizes can resolve the denser regions of molecular clouds, and low-luminosity widespread emission is under the detection limit of the interferometric observations. 

Recently, \citet{2025A&A...703A.176Z} applied a column density fitting method based on Orion B observations to estimate the PDF in a region of the spiral galaxy M51. While our approach is similar, we did not employ sophisticated fitting algorithms to derive parameters such as $N_0$, $\alpha$, and $\sigma_{\eta,10}$. They also estimated $N_{\mathrm{thres}}$, the threshold column density where the log-normal distribution transitions to a power-law tail. In the spiral arm region of M51, the derived parameters were $N_0 \sim 10^{22}\,$\sqc, $\sigma_{\eta,e} \sim 0.2$ ($\sigma_{\eta,10} \sim 0.09$), and $\alpha \sim -2$ (see footnote for differences in notation\footnote{Please note that they define the PDF in linear scale, while we define it in logarithmic scale. They also define $\alpha$ as a positive quantity. This introduces a shift of 1 in the absolute value of the power-law slope and a sign difference (e.g., their $\alpha = 3$ corresponds to our $\alpha = -2$).}). In our observations, the modeled values and the total column density fit the observed data best around log$N_0=21.5$--22.5. Therefore, our observed area seems to have a similar column density as their test region in M51. Meanwhile, we cannot determine whether our models show similar values of $\alpha$ or $\sigma_{\eta,10}$ because the intensity dependence on these parameters is not strong in our model at this value of $N_0$.

\subsection{Comparisons with galactic centers} \label{sec:gal_cen}
Central molecular zones in galactic centers typically have higher volume densities. One might expect the \nthp/HCN ratios to be higher than those in disk regions; however, this is not the case. Single-dish observations of galactic centers using the IRAM 30-m telescope by \citet{2015A&A...579A.101A} observed with the beam sizes of 420\,pc to 8.5\,kpc in linear scale and showed the \nthp/HCN ratios of $0.11 \pm 0.005$ and $0.14 \pm 0.008$ in the star-forming galaxies M83 and M51, respectively; $0.16 \pm 0.003$ in the starburst galaxy NGC~253; $0.07 \pm 0.002$ in M82; $0.08 \pm 0.004$ in the Seyfert galaxy NGC~1068; and $0.21 \pm 0.04$ in the ultra-luminous infrared galaxy Arp~220. Our models reproduce these ratios only at a low column-density regime ($N < 10^{21.5}$\,\sqc).

While some of the low ratios may be influenced by AGN activities due to the overluminous HCN \citep{2015ApJ...799...26M}, even galaxies with minimal AGN contribution (e.g., NGC~253, M82) show ratios comparable to those found in AGN-containing galactic centers. A possible explanation for this trend is the warm temperature in galactic centers. It is known that \nthp\ is enhanced in cold environments where CO is frozen onto dust grains. In galactic centers, CO likely stays in the gas phase and suppresses \nthp\ abundances.

\begin{figure*}
\centering{
\includegraphics[width=0.52\textwidth]{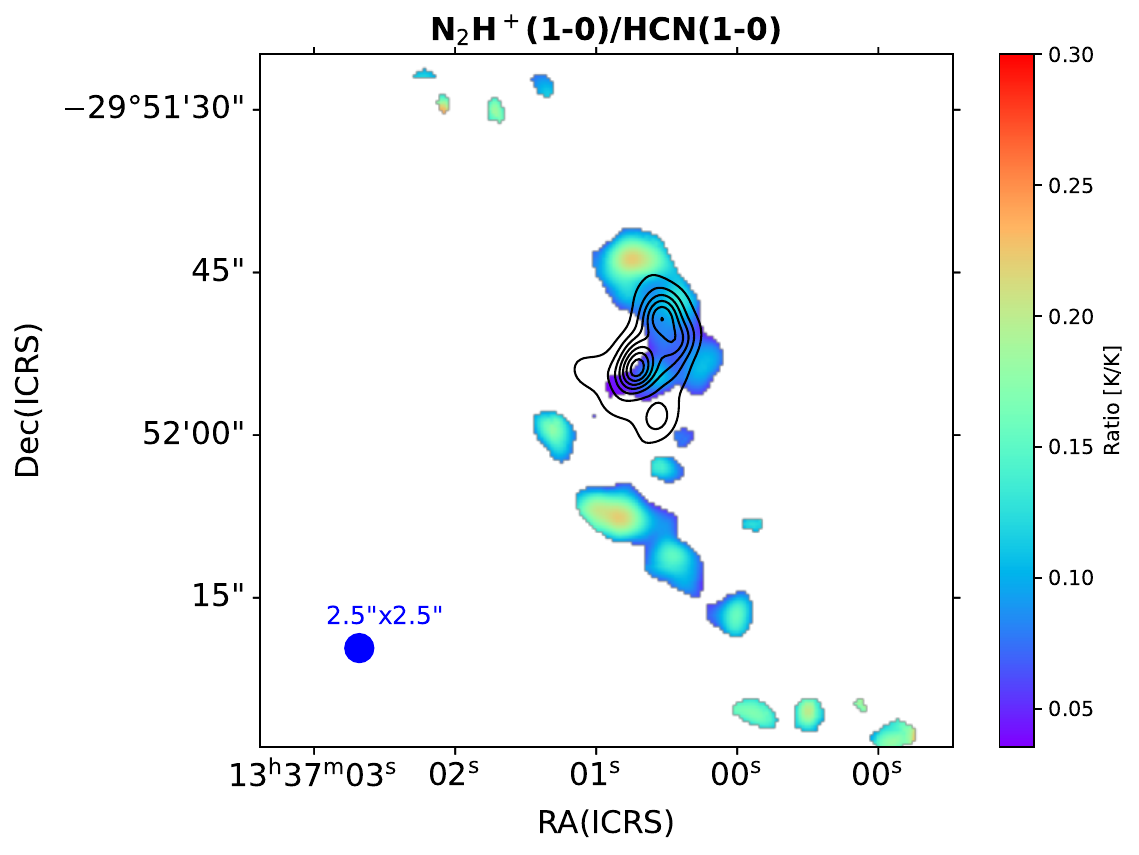}
\includegraphics[width=0.45\textwidth]{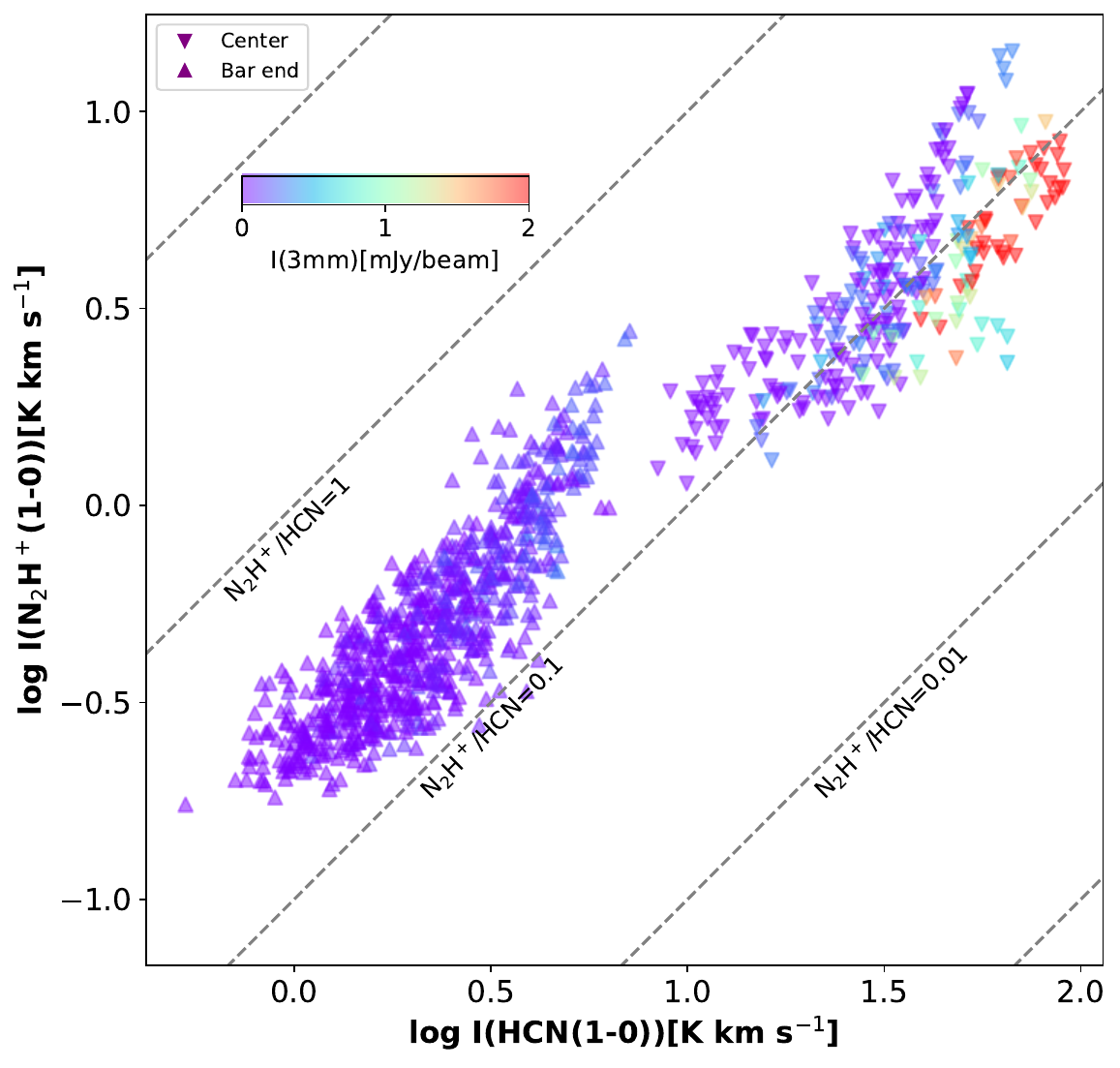}
}
\caption{(Left) The \nthp/HCN ratio toward the central region of M83 shown in the color scale. The black contours show 3-mm continuum intensities with 0.5 m\Jypbm\ increments. The beam size is shown at the bottom left corner. (Right) The same figure as Figure \ref{fig:scatter}(a), but including data points from the galactic center region. The grey dashed lines have slopes of 1, and show the lines of constant \nthp/HCN ratios. \label{fig:center_ratio}}
\end{figure*}

Interferometric observations can help clarify the physical origin of the low \nthp/HCN ratio. To compare the central region and the bar-end region, we used the data towards the central region of M83 presented in \citet{2019ApJ...884..100H}. Its field of view is plotted in Figure \ref{fig:mom0}(f), which is a single-pointing of a beam size $\sim63''$. The procedure to create a mask is similar to the one used for data towards the bar-end region. The common beam size between the HCN and \nthp\ data is $2.5''$. We convolved the HCN cube to twice the size, and masked out the region below 2$\sigma$. This mask was applied to both the HCN and \nthp\ cubes before the velocity-integrated images were created. For the velocity-integrated images, we applied the cutoff of $15\sigma$ and $5\sigma$ of the velocity-integrated images for HCN and \nthp\, respectively, to exclude noisy pixels. The rms noise levels are 20\,mK for HCN and 8.5\,mK for \nthp\ over 10\,\kms. To compare two regions in Figure \ref{fig:center_ratio} right, we convolved the bar-end data to $2.5''$ from $2.2''$. The original pixel size of the central region is $0.3''$, which is regridded for the scatter plot to $0.9''$, so that it avoids oversampling and matches the bar-end region.

It has already been shown by \citet{2019ApJ...884..100H} in ALMA observations towards the central region of M83 that \nthp(1--0) does not peak at the locations of dust continuum or free-free emission, but instead at a region where gas accumulates due to cloud collisions. The \nthp/HCN ratio is generally low (0.03--0.25; Figure \ref{fig:center_ratio} left), with the lowest values found around the optical nucleus of M83. While lower values are partly due to the lower detection limit of ratios compared to the bar-end regions (Figure \ref{fig:center_ratio} right), lower ratios are a general trend in the central region. Because of a relatively constant \hcop/HCN(1--0) ratio in this region compared to the bar-end region \citep{2024ApJ...969...82H}, it is unlikely to result from over-luminous HCN.

While there appears to be an anti-correlation between the free-free emission and the \nthp/HCN ratio as seen in Figure \ref{fig:center_ratio}, a similar suppression of the \nthp/HCN ratio is also seen in regions with strong non-thermal radio emission \citep[synchrotron;][]{1994ApJ...421..122T}. Energetic processes such as star formation and supernova explosions are likely heating dust grains and thereby reducing the \nthp/HCN ratio, consistent with findings in Galactic star-forming regions \citep{2023A&A...679A.112T}.

Not all galactic centers exhibit low \nthp/HCN ratios. Line surveys of the Galactic center region \citep{2012MNRAS.419.2961J} reveal regions within the Sgr~A and Sgr~B clouds where \nthp/HCN~$\gtrsim 0.5$ extending over $\sim 10-20\,$pc. In the Galactic center, star formation is not as active as in the galaxies studied by \citet{2015A&A...579A.101A}, suggesting that the gas and dust may be colder. Spectral energy distribution fitting shows cold dust components with temperatures around $\sim 15\,$K in all the clouds observed by \citet{2004A&A...427..217R} towards the Galactic center, although warmer components ($\sim 30\,$K) are also present.

The results discussed above indicate that \nthp\ is not a reliable tracer of dense gas in galactic centers unless the dust is cold. To accurately probe dense gas in these regions, alternative tracers may be required, such as higher-$J$ transitions of HCN or \hcop, or molecules like HC$_3$N.

\section{Summary}\label{sec:summary}
In this paper, we presented 48-pc resolution observations of the dense gas tracers \nthp\ and HCN, and compared them with the 1-mm and 3-mm continua and archival CO data in the disk region of the spiral galaxy M83. We have also compared the observational results with radiative transfer modeling, considering the column density probability distribution function within the observational beam. Below is a summary of our main findings.

\begin{itemize}

  \item The observed relationship of \nthp\ intensity as a function of the HCN intensity is super-linear with the S/N weighted fit, which is more pronounced in the two-slope model (Figure \ref{fig:scatter}a). This trend is consistent with previous findings in M51, where a largely linear relationship transitions to super-linear behavior at the high-luminosity end. A similar super-linear trend is observed in the \nthp--1\,mm (dust continuum) relationship (Figure \ref{fig:scatter}b), which closely resembles that of \nthp--HCN.

    \item The observed trend can be well reproduced by radiative transfer models that incorporate column-density-dependent line intensities from Galactic observations and the column density probability distribution function (Figure \ref{fig:model}). This result implies that extragalactic clouds in the observed regions of M83 can be expressed as combinations of Galactic clouds. The log-normal-only model provides a steeper super-linear \nthp-HCN relationship than the log-normal+power-law model.

    \item The radiative transfer modeling demonstrates that \nthp\ emission is most sensitive to the presence of a power-law tail, whereas the intensities of other transitions are less affected (Figure \ref{fig:model_slope}). However, the difference caused by the slope of the power-law is still insignificant when the peak of the column-density log-normal distribution is high ($N_{\mathrm{peak}} \gtrsim 10^{21.5}\,$\sqc).

    \item The HCN--1\,mm (dust) relationship resembles the \nthp--HCN relationship but exhibits less pronounced super-linearity (Figure \ref{fig:scatter}d). This near-linear trend is well reproduced by the radiative transfer model. The HCN intensities vary only modestly (by $\lesssim 0.5\,$dex) with the addition of a power-law tail, provided that $N_{\mathrm{peak}} \geq 10^{21}\,$\sqc.

    \item Relationships between CO intensities and those of HCN, \nthp, and the 1\,mm continuum show weaker correlations compared to the relationships among HCN, \nthp, and the 1\,mm continuum (Figure \ref{fig:scatter_co}). This likely reflects the fact that CO traces less dense gas and becomes optically thick even at moderate column densities. This trend is consistent with the results from the radiative transfer modeling.

    \item The 3\,mm continuum, which traces the star formation rate, also shows weaker correlations with other quantities. This can be explained by the breakdown of the Kennicutt-Schmidt law on spatial scales smaller than 100\,pc, as previously observed in Local Group galaxies.

    \item The archival data toward the center of M83 indicate that \nthp\ is not a reliable tracer of dense gas in galactic center regions if the dust temperature is high (Figure \ref{fig:center_ratio}).

\end{itemize}

\begin{acknowledgments}
We thank the anonymous referee for the careful reading of this manuscript. We are also grateful to Dr. Kazushi Sakamoto for the helpful comments on this manuscript. This paper makes use of the following ALMA data: \\
ADS/JAO.ALMA\#2016.1.00164.S\\
ADS/JAO.ALMA\#2015.1.01177.S\\
ADS/JAO.ALMA\#2021.1.01195.S.\\
ALMA is a partnership of ESO (representing its member states), NSF (USA) and NINS (Japan), together with NRC (Canada), MOST and ASIAA (Taiwan), and KASI (Republic of Korea), in cooperation with the Republic of Chile. The Joint ALMA Observatory is operated by ESO, AUI/NRAO and NAOJ. Data analysis was carried out on the Multi-wavelength Data Analysis System operated by the Astronomy Data Center (ADC), National Astronomical Observatory of Japan. N.H. acknowledges support from JSPS KAKENHI grant No. JP25K07375, and was supported by the ALMA Japan Research Grant of NAOJ ALMA Project, NAOJ-ALMA-400. J.K. acknowledges support from NSF through grants AST-2006600 and AST-2406608, and also from STScI through grants HST-GO-17747.001-A and JWST-GO-12484.001-A. S.K.S is supported by an International Research Fellowship of the Japan Society for the Promotion of Science (JSPS). Y.N. gratefully acknowledges support from JSPS KAKENHI Grant Numbers JP23K13140 and JP23K20035. Grammatical errors in this manuscript were corrected using Grammarly\footnote{https://www.grammarly.com}.
\end{acknowledgments}

%

\vspace{5mm}
\facilities{ALMA}


\software{astropy \citep{2013A&A...558A..33A,2018AJ....156..123A}, CASA \citep{2007ASPC..376..127M, 2022PASP..134k4501C} 
          }



\appendix

\section{Physical quantities derived from the measurements}\label{app:phys}

Quantities measured in observations, CO and HCN integrated intensities, and 1-mm and 3-mm continuum intensities can be used to estimate physical parameters using the following formulae. 

The overall molecular gas surface density is calculated using a typical conversion factor of $\alpha_{CO}=4.3\,M_{\odot}$\,pc$^{-2}$(K\,\kms\,)$^{-1}$ \citep{2013ARA&A..51..207B}. 
\begin{eqnarray}
\Sigma_{\rm mol} = \alpha_\mathrm{CO} \int T_b\,dv  \times cos(i) \nonumber \\
= 3.9\,  \mathrm{({\it M}_{\odot}\,pc^{-2}) \int T_b\,dv (K\,km\,s^{-1})^{-1}},
\end{eqnarray}
where the velocity-integrated intensity is described as $\int T_b\,dv$, and $i$ is the inclination angle. We use $i=26^{\circ}$ reported by \citet{2023ApJ...949..108K} (cos($i$)=0.9). 

The dense gas surface density is calculated as
\begin{eqnarray}
\Sigma_{\rm dense} = \alpha_\mathrm{HCN} \int T_b\,dv  \times cos(i) \nonumber \\
= 9.0\,  \mathrm{({\it M}_{\odot}\,pc^{-2}) \int T_b\,dv (K\,km\,s^{-1})^{-1}}.
\end{eqnarray}
We use the conversion factor $\alpha_{\rm HCN}= 10\, M_{\odot}$\,pc$^{-2}$ (K\,km\,s$^{-1}$)$^{-1}$ \citep{2004ApJS..152...63G}. 

The star formation rate $\dot{M}_*$ within the aperture $A$ is \citep[e.g., ][]{2016MNRAS.463..252B}.
\begin{eqnarray}
\dot{M}_*/A= 9.49 \times 10^{10} \,M_{\odot} \,{\rm yr}^{-1} \,\mathrm{pc}^{-2} \,g_{\rm ff}^{-1}\left(\frac{\alpha_B}{\rm cm^3\,s^{-1}}\right) \nonumber\\ \times \left(\frac{T_\mathrm{e}}{\mathrm{ K}}\right)^{0.5}  \left(\frac{D}{\rm Mpc}\right)^2 \left(\frac{f}{\rm Jy\,beam^{-1}}\right) \left(\frac{A}{\mathrm{pc^2\,beam^{-1}}}\right)^{-1},
\end{eqnarray}
where $g_{\rm ff}$ is the Gaunt factor, $\alpha_B$ is the effective recombination coefficient, $T_{\rm e}$ is the electron temperature that we assume to be 5000\,K, $D$ is the distance to the galaxy, $f$ is the 3-mm flux per beam, and $A$ is the beam size area in pc. The Gaunt factor is shown as
\begin{equation}
g_{\rm ff} = 0.5535\, {\rm ln} \left[\left(\frac{T_{\rm e}}{\rm \,K}\right)^{1.5}\left(\frac{\nu}{\rm GHz}\right)^{-1}Z^{-1}\right]-1.682,
\end{equation}
where $\nu$ is the frequency and $Z$ is the electric charge, which can be approximated as $\sim 1$.
We use the effective recombination coefficient $\alpha_B$ \citep{2020ApJ...895...85M}
\begin{equation}
    \alpha_\mathrm{B} = 3.63 \times 10^{-10} \left(\frac{T_{\rm e}}{\rm K}\right)^{-0.79}.
\end{equation}
The star formation rate surface density can be obtained by
\begin{equation}
\Sigma_\mathrm{SFR} (M_{\odot} \,{\rm yr}^{-1} {\rm pc}^{-2})= \dot{M}_*/A \times cos(i),
\end{equation}
where $A$ is the aperture size in pc$^{2}$.

We derive the molecular column density using the method proposed by \citet{1983QJRAS..24..267H} and simplified by \citet{2019ApJ...871..170M}, also considering the inclination.
\begin{equation}\label{eq:hildebrand}
 N(\mathrm{H_2}) = 7\times 10^{22} \mathrm{cm}^{-2} \,cos(i)\,R_{gd} \left(\frac{\lambda (\mathrm{mm})}{0.4}\right)^{\beta} \left(\frac{T_R(\mathrm{K})}{T_d(\mathrm{K})}\right),   
\end{equation}
where $R_{gd}$ is the gas-to-dust ratio, $\lambda$ is the wavelength, $\beta$ is the emissivity, $T_R$ is the brightness temperature of the dust continuum ($\lambda\sim1$\,mm in our case), and $T_d$ is the dust temperature. We use $\beta=2.1$, a value derived by \citet{2012MNRAS.421.2917F} around the galactocentric radius of $R=3-4\,$kpc, which corresponds to our observed region. We set $T_d=20\,$K because the dust temperature obtained from Herschel is 20--25\,K \citep{2012MNRAS.421.2917F}. According to \citet{2024ApJ...968...97L}, $R_{gd}\sim 100-150$, for $R=3-4\,$kpc, and we use the value of $R_{gd}=125$. This estimation is subject to uncertainty of a factor of a few from the dust opacity per unit dust $\kappa_{\nu}$. \citet{1994A&A...291..943O} derived a value $\sim 1\,$cm$^2$\,g$^{-1}$ at a wavelength of $\lambda\sim1\,$mm, while $\kappa_{\nu}=0.294\,$cm$^2$\,g$^{-1}$ according to \citet{2003ARA&A..41..241D}.

\section{Column density dependence of molecular abundances and physical parameters}\label{app:chem}
To model the molecular emission at certain column densities, we adopt the parameters derived by \citet{2021A&A...646A..97T} as a result of fitting to their observations in the Galactic molecular cloud Perseus. For the volume density, we used
\begin{equation}\label{eq:dens}
    n = 2\times 10^4 \,\mathrm{cm^{-3}}\left(\frac{N}{10^{22}\,\mathrm{cm}^{-2}}\right)^{0.75}, 
\end{equation}
where $n$ is the volume density and $N$ is the column density of H$_2$.

\citet{2021A&A...646A..97T} derived analytical formulae for column-density-dependent molecular fractional abundances based on their Galactic observations, and all of the expressions below are taken from their fit. The fractional abundances are expressed as
\begin{equation}
    X(N) = X_0 \times f_\mathrm{out}(N) \times f_\mathrm{in}(N),
\end{equation}
where $X_0$ is a constant for scaling, whose value for each species is CO: $9.5\times 10^{-5}$, HCN: $3\times 10^{-9}$, \nthp: $1.5\times 10^{-10}$. The factor $f_\mathrm{out}$ accounts for effects of PDRs at the outer part of molecular clouds and has the form
\begin{equation}
    f_\mathrm{out}(N) = 
    \begin{cases}
        [e^{(A_V-A_0)}]^6 &\text{if $A_V \leq A_0$}\\
        1 &\text{if $A_V > A_0$,}
    \end{cases}
\end{equation}
where $A_0=2$\,mag. is used for CO, HCN, and \nthp. We use the conversion factor between the column density and the visual extinction as $N(\mathrm{H}_2)/A_V = 9.35\times 10^{20}\,$\sqc. The factor $f_\mathrm{in}$ corrects for the freeze-out in the dense inner clouds. For HCN and CO, the factor is
\begin{equation}
    f_\mathrm{in}(N) = e^{-n/n_\mathrm{fr}}
\end{equation}
with $n_\mathrm{fr}=10^5$\,cm$^{-3}$ for CO and $n_\mathrm{fr}=2\times 10^5$\,cm$^{-3}$ for HCN. A different expression is used for \nthp, because this species is somewhat freeze-out resistant:
\begin{equation}
      f_\mathrm{in}(N) = 1- e^{-(n/n_\mathrm{fr})^2} 
\end{equation}
with $n_\mathrm{fr}=3\times 10^4$\,cm$^{-3}$.

\begin{figure}
\includegraphics[width=0.8\textwidth]{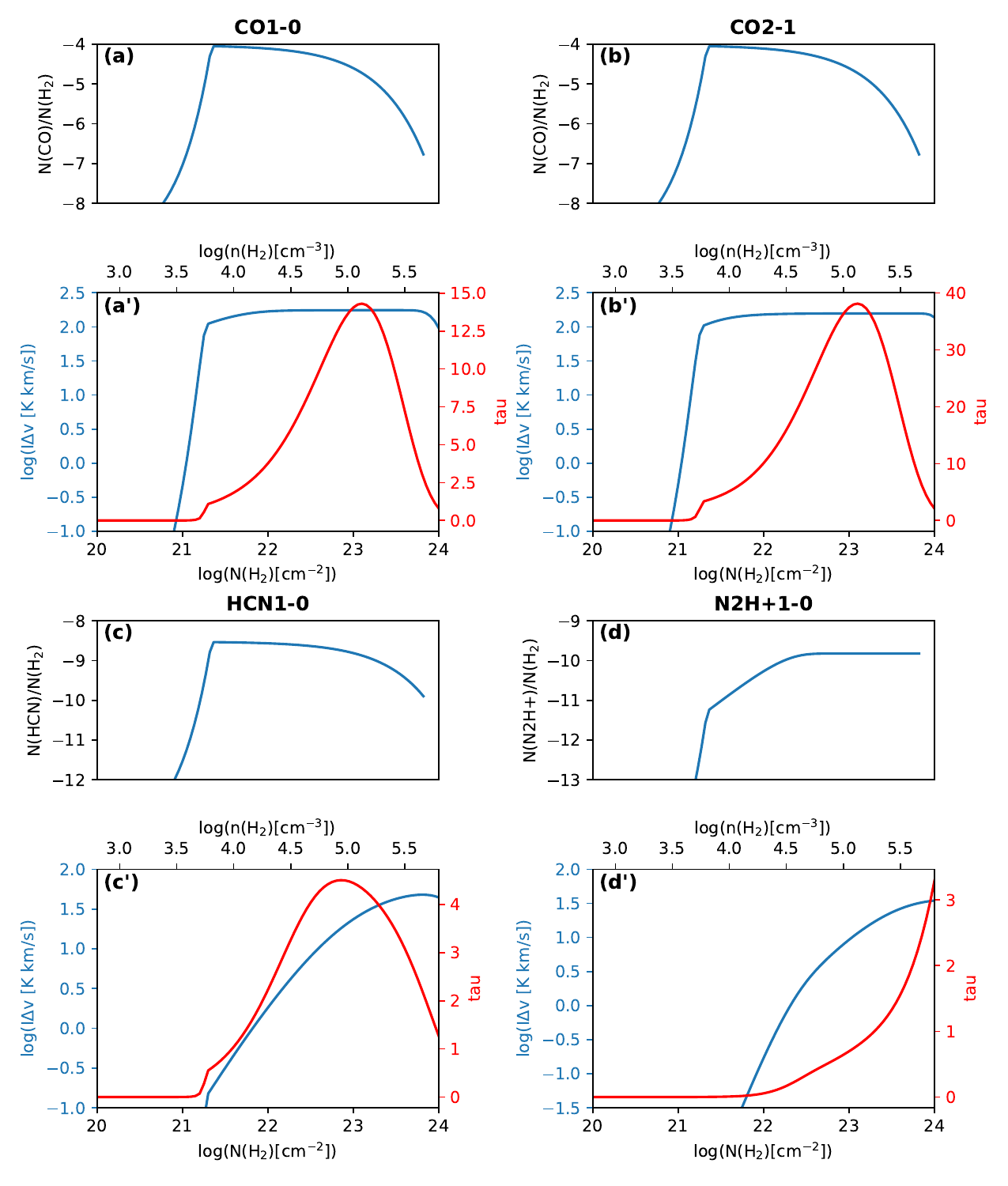}
\caption{Column-density dependent fractional abundances by \citet{2021A&A...646A..97T} (Panels a-d), simulated integrated intensities from these abundances (Panels a'-d', blue) and optical depth (Panels a'-d', red) by RADEX.\label{fig:tafalla_model}}
\end{figure}

Figure \ref{fig:tafalla_model} shows integrated intensities as functions of column densities based on the RADEX calculation from the above fractional abundance model, as well as their corresponding optical depths. Because we use $T=20\,$K for CO, CO(1--0) is optically thinner than CO(2--1), although it becomes optically thick with a relatively low total column density. HCN is less optically thick than CO transitions, but it becomes optically thick already around $N=10^{21}\,$\sqc. Meanwhile, \nthp\ is optically thin until $N\sim 10^{23}\,$\sqc. 

\section{Column-density distribution}\label{app:logn}
Turbulent theory of star formation says that the distributions of volume densities have log-normal shapes \citep{2002ApJ...576..870P,2005ApJ...630..250K} with the following expression:
\begin{equation}
    p(x)dx = \frac{1}{\sqrt{2\pi \sigma_{s,e}^2}}\mathrm{exp}\left[-\frac{1}{2}\left(\frac{x-\overline{x}}{\sigma_{s,e}}\right)^2\right] dx,
\end{equation}
where $x$ is the logarithm of the volume density $n$ ($x$=ln\,$n$), $\overline{x}$ is the mean of the logarithm of volume density ($\bar{x} = \overline{\mathrm{ln}\,n}$), $\sigma_{s,e}$ is the standard deviation of $x$. The meaning of the subscript of $\sigma$ is explained later in this section. We can express $\overline{x}$ in terms of $\sigma$ as follows
\begin{equation}
    \overline{x}=-\frac{\sigma_{s,e}^2}{2}.
\end{equation}
Here, $\sigma_{s,e}$ can be obtained from the rms Mach number of the flow $\mathscr{M}$
\begin{equation}
    \sigma_{s,e}^2  =\mathrm{ln}(1+\gamma^2 \mathscr{M}^2)
\end{equation}
where $\gamma\sim0.5$ is a parameter determined from simulations. In addition to the log-normal term, it has been observationally suggested that there is a power-law term at the higher-density end ($p(x) \propto n^{\alpha}$) if the molecular gas is about to gravitationally collapse to form stars.

While it is easier to consider volume densities in theoretical studies, observable quantities are column densities, i.e., integrated volume densities in the line of sight. We denote the dispersion in the column density $\sigma_{\eta}$, while the dispersion in the volume density is denoted as $\sigma_s$. The distribution of column densities is also found to have log-normal-like shapes with power-law tails. Although some studies found deviations from the log-normal + power-law shapes, such as double-peak distributions, we chose log-normal(+power-law) distributions in our model for simplicity. There are two notations used in the literature; some use natural logarithms while others use logarithms with base 10, which causes a factor of ln10=2.3 difference in the dispersion. We use the subscript 10 or $e$ to describe the dispersion to specify which base is used for the logarithm (e.g., $\sigma_{s,10}$ or $\sigma_{s,e}$). According to \citet{2013ApJ...766L..17S}, the ratio of the dispersion $\sigma$ in the two-dimension and the three-dimension ($\sigma_s/\sigma_{\eta}$) is about 2.5.  Observations of CO(1--0) towards M51 by \citet{2013ApJ...779...44H} found $\sigma_{\eta,10} \sim 0.3-0.5$, which should translate to $\sigma_{s,10} = 0.7-1.2$ assuming $\sigma_s/\sigma_{\eta}=2.5.$ Because a typical line width (FWHM) obtained in our observations is 20\,km\,s$^{-1}$ (the velocity dispersion $\Delta v=8.5\,$km\,s$^{-1}$) and the sound speed at 10\,K in the interstellar medium is $c_s=0.2\,$km\,s$^{-1}$, $\mathscr{M}\sim 40$. This value should translate to $\sigma_{\eta,10}=0.37$. Therefore, we use $\sigma_{\eta,10}=0.4$ as our fiducial value.

The power-law term comes from gravitationally contracting clouds. They have the density distribution as a function of radius \(n(r)\propto r^{-k}\), where $n$ is the volume density and $r$ is the radius \citep{1977ApJ...214..488S}. According to the theoretical study by \citet{2013ApJ...763...51F}, \(p(logN)\propto N^{\frac{2}{1-k}}\). If we express the power-law as \(p(logN)\propto N^{-\alpha}\) and use $k=1.5-2.5$ from \citet{2011A&A...529L...6A}, $\alpha=-4$ to $-1.3$. Observed values reported by \citet{2022A&A...666A.165S} range from $-4.5$ to $-1$, with the median values close to $-2.5$. The actual column density distributions are not simple single power-laws, as some distributions have power-law tails with two different slopes in the PDF of the same cloud. In this paper, we use the slope of $\alpha=-1.5$ as a fiducial value, and discuss the effects of variation. This fiducial value is a relatively flat slope that enhances the higher density term. We choose this value to highlight the difference between the log-normal-only shape and the one with the power-law tail. We change from the log-normal shape to the power-law one when the gradient of the log-normal shape becomes the same as that of the power-law slope.

\section{Additional exploration of parameter variations}\label{app:plots}

\begin{figure*}
\centering{
\includegraphics[width=0.9\textwidth]{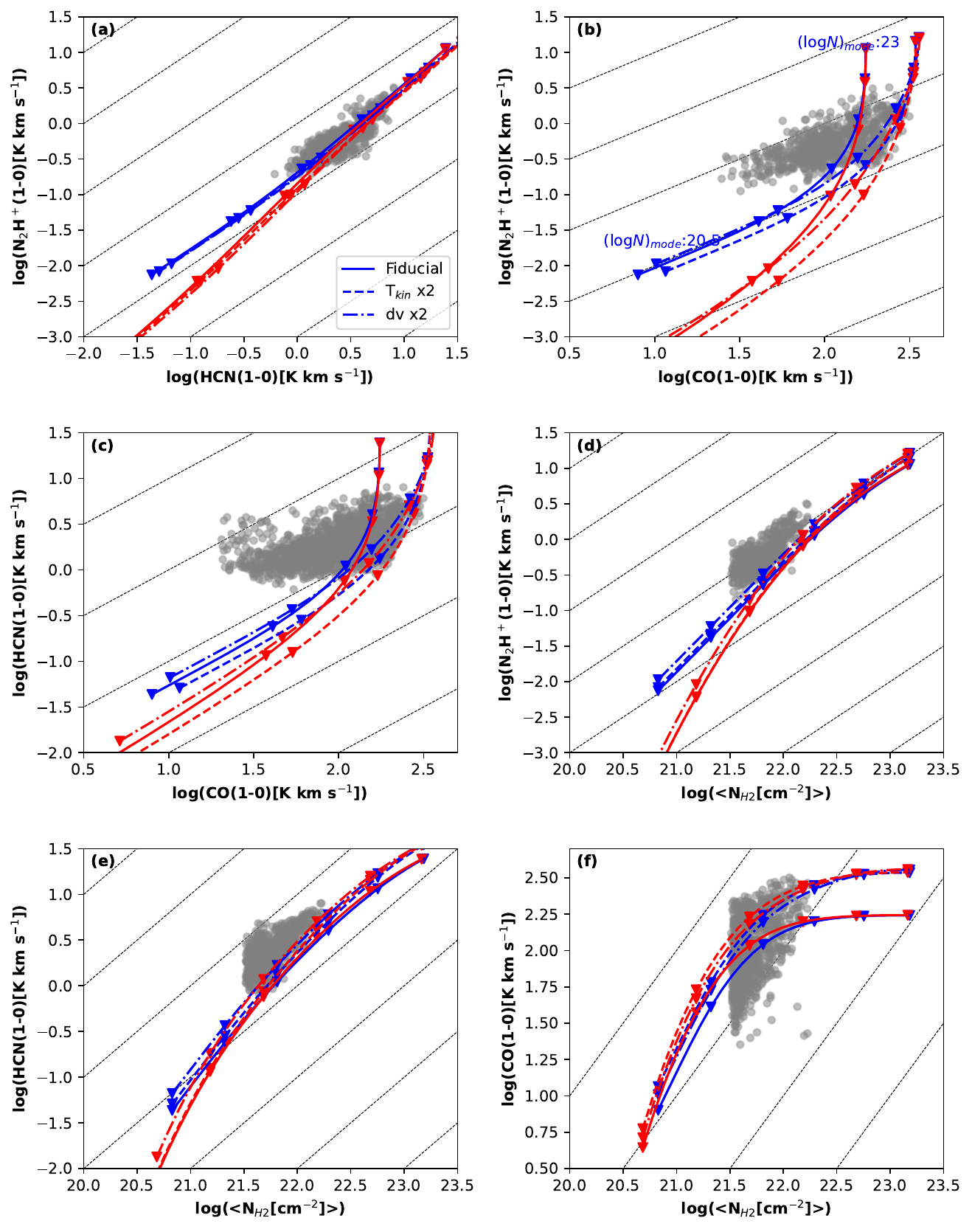}
}
\caption{Modeled intensities of the fiducial model compared with cases where the temperature is doubled, or the velocity width is doubled. Red lines represent models with log-normal N-PDF and blue lines represent models with log-normal plus power-law N-PDF.\label{fig:model_Tdv}}
\end{figure*}

In this section, we further explore the model introduced in Section \ref{sec:rad} by varying model parameters. Figure \ref{fig:model_Tdv} shows the variation of emission intensities when the kinetic temperature is doubled and when the line width is doubled compared to the fiducial case of $T_\textrm{kin}$=20\,K for CO and HCN, $T_\textrm{kin}=10$\,K for \nthp, $\Delta v$=10\,\kms\ for CO, and $\Delta v$=5\,\kms\ for HCN and \nthp. As expected, increases in the temperature or line width have more impact on an optically thick transition such as CO. The difference is about a factor of 2 when $N_\textrm{peak}$ is high. Even for CO, the difference is less for lower $N_\textrm{peak}$ due to the lower optical depth.

\begin{figure*}
\centering{
\includegraphics[width=0.9\textwidth]{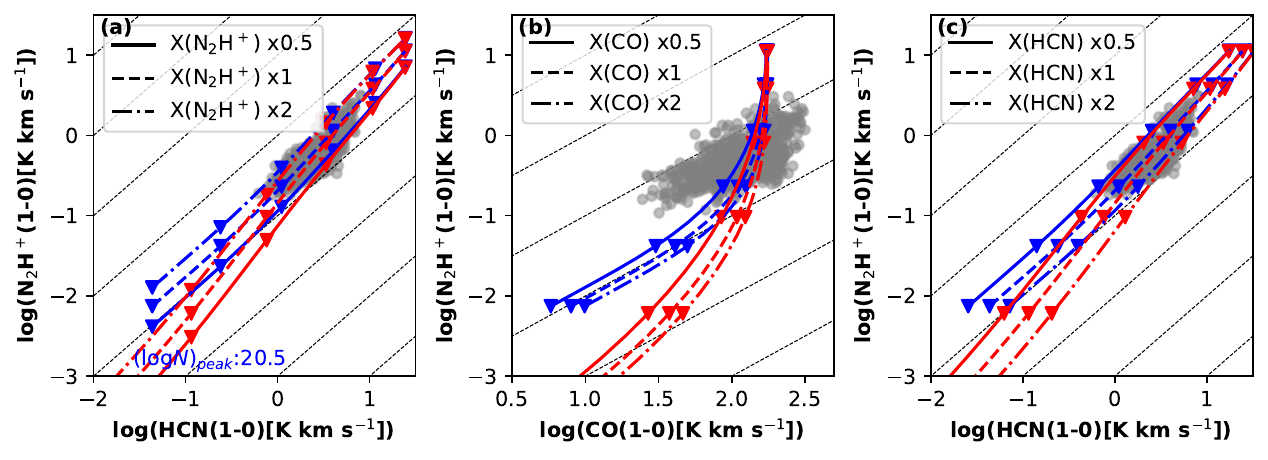}
}
\caption{Modeled intensities with varying fractional abundances of \nthp(left), CO(middle), and HCN(right).\label{fig:model_ab}}
\end{figure*}

Variations of fractional abundances affect resultant intensities more in optically thin cases, as one would imagine from the basic principles of radiative transfer (Figure \ref{fig:model_ab}). Changes in fractional abundances of \nthp\ and HCN vary their intensities almost linearly for low column densities, while changes become less for higher column densities. The intensities of CO do not vary with changes in the fractional abundance of CO at high column densities because the optical depth is very high.


\bibliography{sample631}{}
\bibliographystyle{aasjournalv7.1}



\end{document}